\documentclass{article}
\usepackage[pdftex]{graphicx}
\usepackage{spconf,amsmath}
\usepackage{booktabs}
\usepackage{cite}
\usepackage[subrefformat=parens]{subcaption}

\title{ANOMALOUS SOUND DETECTION BASED ON\\
INTERPOLATION DEEP NEURAL NETWORK}
\name{Kaori Suefusa, Tomoya Nishida, Harsh Purohit, Ryo Tanabe, Takashi Endo, and Yohei Kawaguchi}
\address{Research and Development Group, Hitachi, Ltd.\\
1-280, Higashi-koigakubo, Kokubunji-shi, Tokyo 185-8601, Japan\\
\texttt{kaori.suefusa.fz@hitachi.com}}
\begin{document}
%
\maketitle
\begin{abstract}
As the labor force decreases, the demand for labor-saving automatic anomalous sound detection technology that conducts
maintenance of industrial equipment has grown.
Conventional approaches detect anomalies
based on the reconstruction errors of an autoencoder.
However, when the target machine sound is non-stationary, a reconstruction error tends to be large independent of an anomaly,
and its variations increased because of the difficulty of predicting the edge frames.
To solve the issue, we propose an approach to anomalous detection
in which the model utilizes multiple frames of a spectrogram
whose center frame is removed as an input,
and it predicts an interpolation of the removed frame as an output.
Rather than predicting the edge frames,
the proposed approach makes the reconstruction error consistent with the anomaly.
Experimental results showed that the proposed approach achieved 27$\%$ improvement
based on the standard AUC score, especially against non-stationary machinery sounds.
\end{abstract}


\section{Introduction}
\label{sec:intro}
All machinery in factories is subject to failures or breakdown,
causing companies to bear significant costs.
Conventionally, skilled maintenance technicians have diagnosed a machine's condition
by listening to the machinery.
However, as the labor force decreases, it has become difficult
to maintain the quality of the maintenance service with fewer skilled workers.
To solve the issue, technology that performs automatic diagnosis based on operating sounds
has been developed \cite{Yamashita2006, Koizumi2017}.

Conventional approaches to unsupervised anomaly detection
employed autoencoders and attempted to detect anomalies
based on reconstruction errors \cite{Chandola2009}.
In terms of anomalous sound detection,
multiple frames of a spectrogram are used as an input feature,
and the same number of frames are generated as an output.
Although such approaches can achieve high performance,
some issues have been remained.
When the target machine sound is non-stationary,
the reconstruction error tends to be large without regarding the anomaly.
Since it is relatively difficult to predict edge frames,
the variation of the reconstruction error can be large.

In this paper, we propose an unsupervised approach
to anomalous sound detection
called the ``interpolation deep neural network (IDNN).''
The model utilizes multiple frames of a spectrogram
whose center frame is removed as an input,
and it predicts an interpolation of the removed frame as an output.
Anomalies can be detected based on an interpolation error
that is the difference between the predicted frame and the true frame.
It is hypothesized that
the proposed IDNN will not be affected
by variations of errors regarding the edge frame,
since it does not predict them.

We experimented
to compare the performance of
our approach with the conventional one
using real-life industrial machine sounds.
Experimental results indicated that
our IDNN outperformed the conventional approach,
especially against non-stationary machinery sounds.

\section{Conventional Approaches}
\label{sec:methods}


\begin{figure}[t]
	\begin{center}
		\begin{tabular}{c}
			\begin{minipage}{0.8\hsize}
			\centering
			\includegraphics[width=\columnwidth, clip]{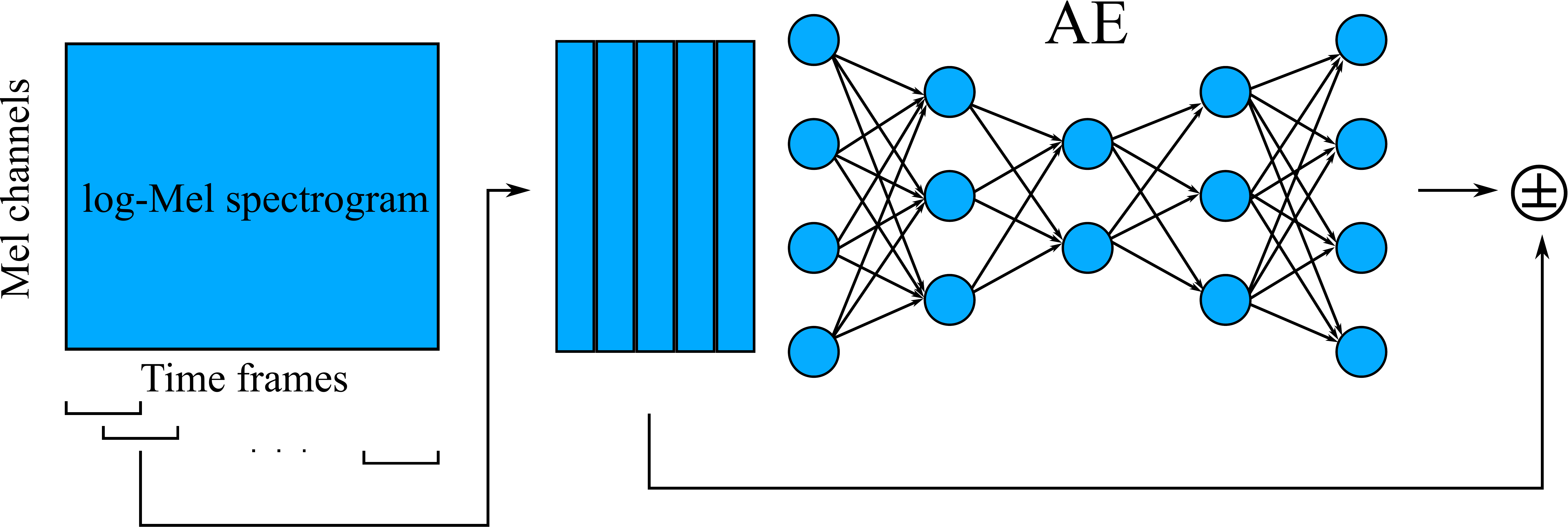}
			\vspace{-1.5em}\subcaption{}\vspace{1em}
			\label{fig:AE}
			\end{minipage}
			\\
			\begin{minipage}{0.8\hsize}
			\centering
			\includegraphics[width=\columnwidth, clip]{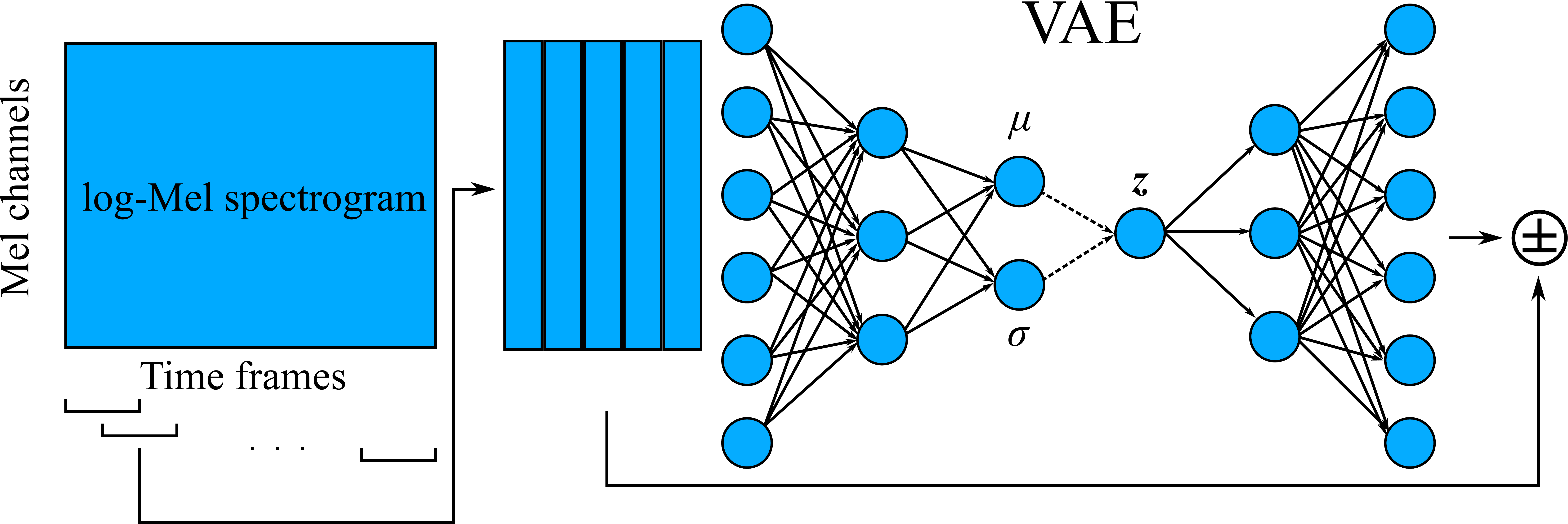}
			\vspace{-1.5em}\subcaption{}\vspace{1em}
			\label{fig:VAE}
			\end{minipage}
		\end{tabular}
		\caption{Typical architecture of (a) AE and (b) VAE}
		\label{fig:ae_vae}
	\end{center}
\end{figure}



Several approaches to implementing unsupervised anomalous sound detection have been proposed.
Recent studies leveraged a deep neural network (DNN)
that includes an autoencoder (AE), a variational autoencoder (VAE),
and so forth \cite{Marchi2015, Koizumi2017, Kawaguchi2017, Tagawa2015, Marchi2015RNN}.
To detect anomalies with an AE \cite{Vincent2010},
the model is trained with normal training data
and learns to minimize reconstruction errors
\cite{Marchi2015, Koizumi2017, Kawaguchi2017}.
A reconstruction error is the difference
between the original input and the reconstructed output.
Since the AE is trained with normal data,
the reconstruction error of the normal data is expected to be small
while that of the anomalies would be relatively large.
Thereby, the anomaly score is calculated as the reconstruction error.
Figure \ref{fig:AE} summarizes the typical architecture of an AE for anomalous sound detection.
Parameters of an encoder E($\cdot$) and a decoder D($\cdot$) of an AE
are trained to minimize the loss function given as follows:
\begin{align}
  L_{\text{AE}}
  = \left\lVert \mathbf{x}- D( E(\mathbf{x}) )
  \right\rVert_2^2
  ~\text{,}
  \label{eq:loss_func_AE}
\end{align}
where $\mathbf{x}$ represents an input.

In a manner similar to an AE,
a VAE \cite{Kingma2014} has been also utilized
for anomalous sound detection \cite{An2015, Kawachi2018}.
Figure \ref{fig:VAE} shows the typical architecture of a VAE.
The loss function of a VAE is given as follows:
\begin{align}
  L_{\text{VAE}}
  = w D_{\text{KL}} \left[q(\mathbf{z}|\mathbf{x})||
    p(\mathbf{z}|\mathbf{x})) \right] 
    - E_{q(\mathbf{z}|\mathbf{x})}\log p(\mathbf{z}|\mathbf{x})
  ~\text{,}
  \label{eq:loss_func_VAE}
\end{align}
where $\mathbf{z}$ represents the latent vector and
$D_{\text{KL}}$ represents the Kullback–Leibler divergence
of the approximate posterior and the prior distribution.

Although conventional approaches can achieve high performance,
the following issues remained.
1) In the case of non-stationary sound,
the reconstruction error tends to be large without regarding the anomaly
due to the difficulty of predicting the edge frame.
2) The number of parameters is relatively large
since those approaches attempt to reconstruct the whole input feature,
which consists of multiple frames.
3) As its prediction includes its input itself,
it can fall into a trivial solution
and cannot embed a spectrotemporal structure of normal sound
if the number of bottleneck neurons is set to a large number.

\section{Proposed Approach}
\begin{figure}[t]
	\begin{center}
		\begin{tabular}{cc}
			\begin{minipage}{0.48\hsize}
			\centering
			\includegraphics[width=\columnwidth, clip]{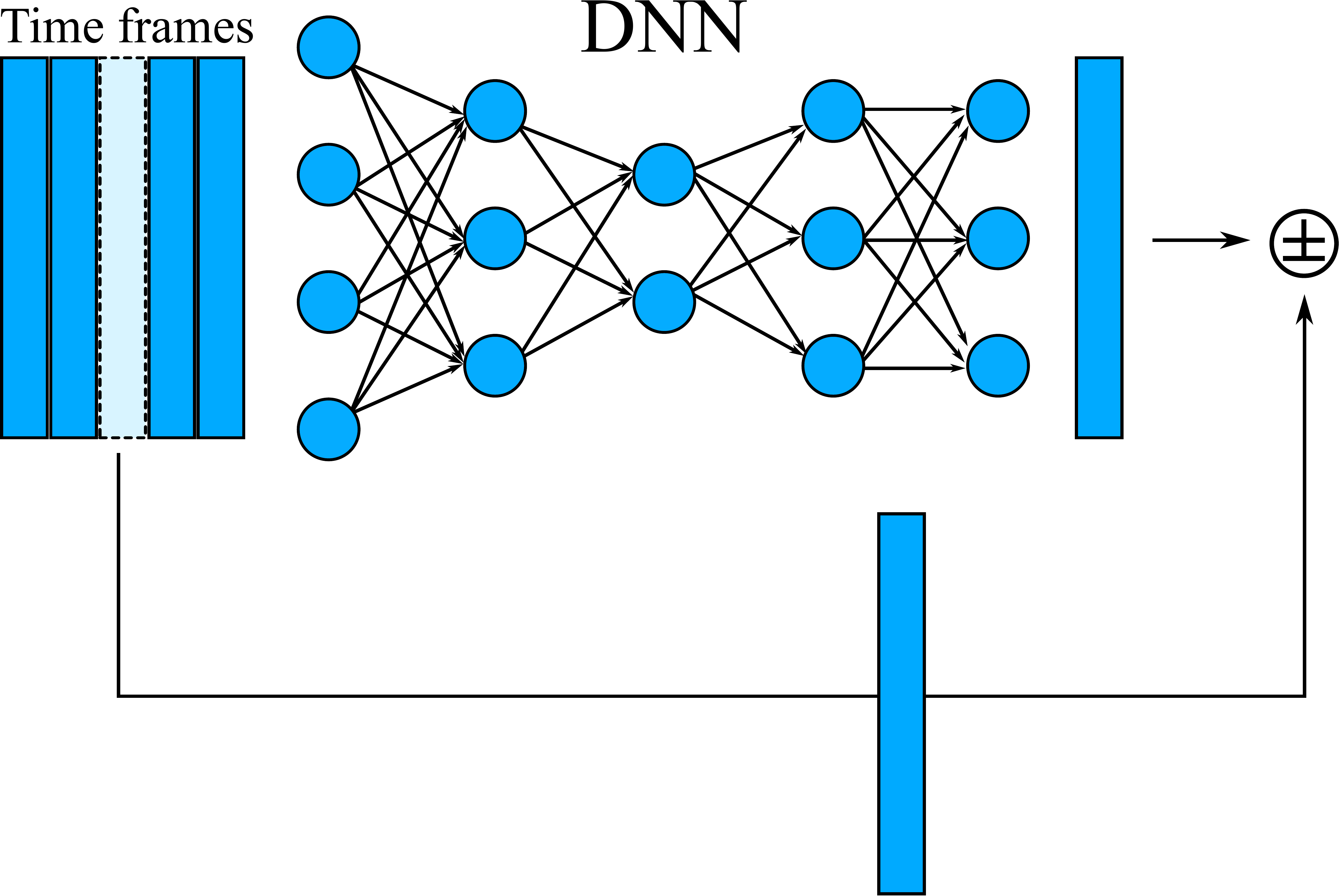}
			\vspace{-1.5em}\subcaption{}\vspace{1em}
			\label{fig:IDNN}
			\end{minipage}
			&
			\begin{minipage}{0.48\hsize}
			\centering
			\includegraphics[width=\columnwidth, clip]{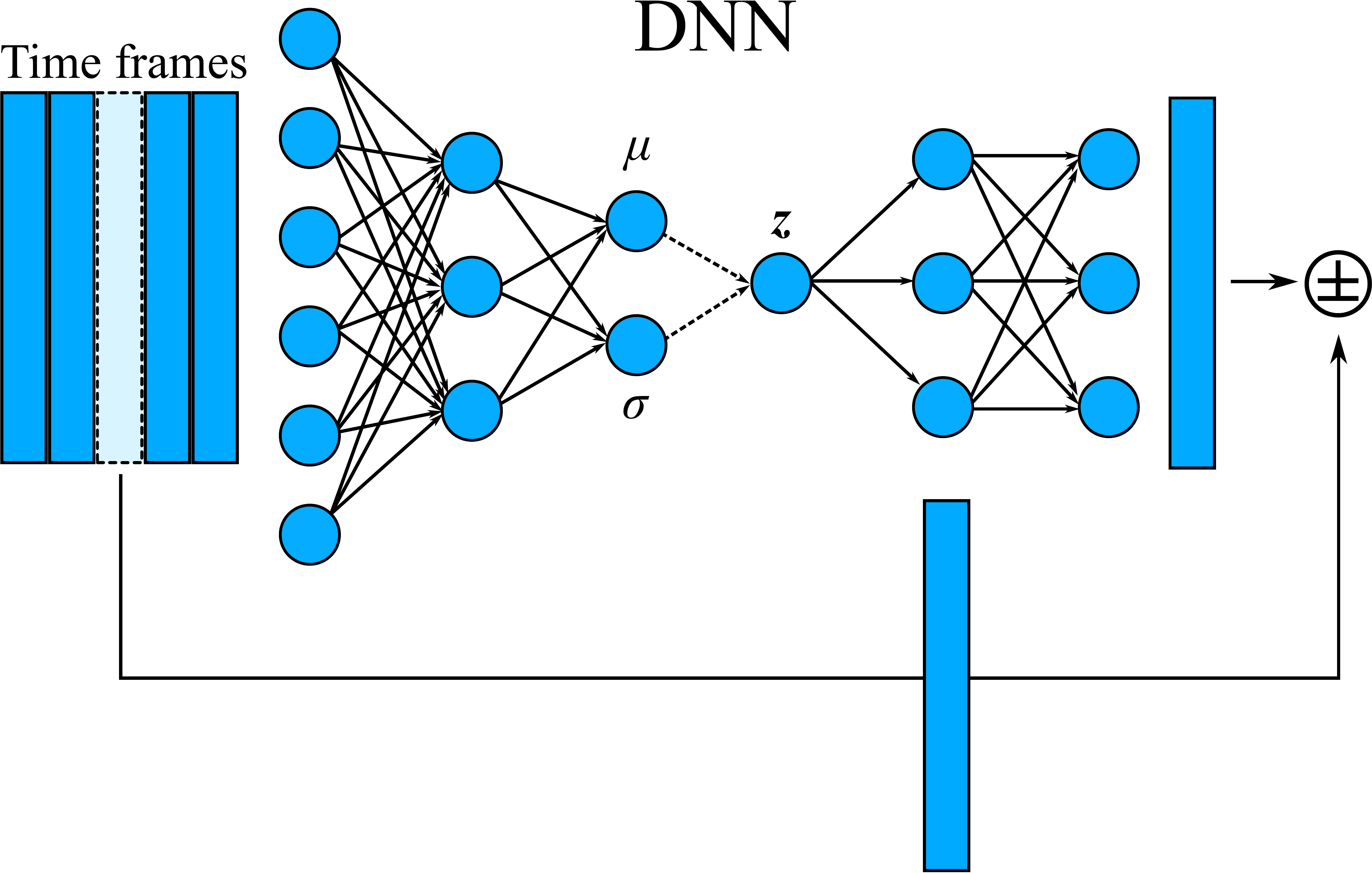}
			\vspace{-1.5em}\subcaption{}\vspace{1em}
			\label{fig:VIDNN}
			\end{minipage}
		\end{tabular}
		\caption{Proposed architecture of (a) IDNN and (b) VIDNN}
		\label{fig:IDNN_VIDNN}
	\end{center}
\end{figure}

\begin{figure}[t]
	\begin{center}
		\begin{tabular}{cc}
			\begin{minipage}{0.48\hsize}
			\centering
			\includegraphics[width=\columnwidth, clip]{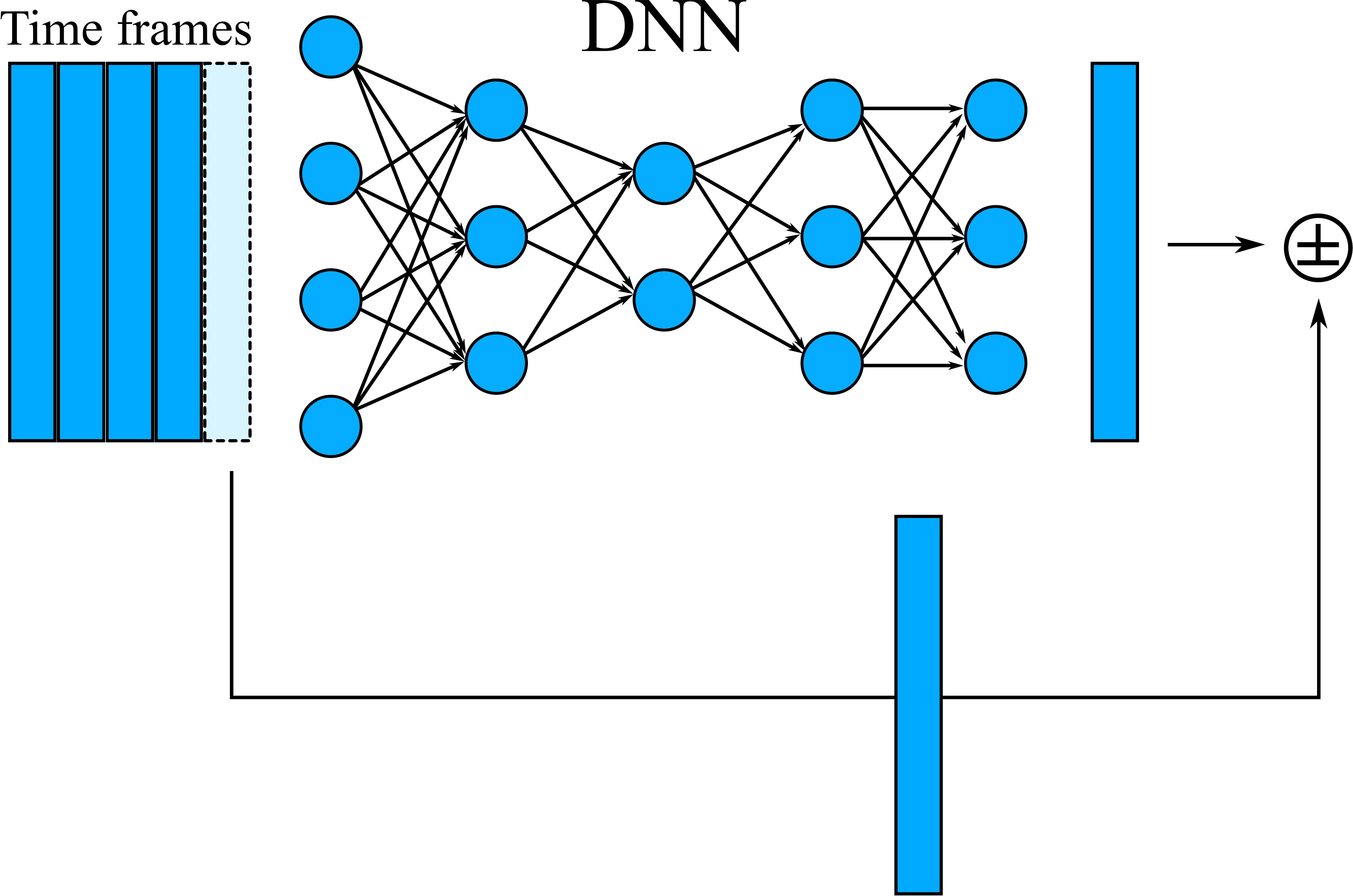}
			\vspace{-1.5em}\subcaption{}\vspace{1em}
			\label{fig:PDNN}
			\end{minipage}
			&
			\begin{minipage}{0.48\hsize}
			\centering
			\includegraphics[width=\columnwidth, clip]{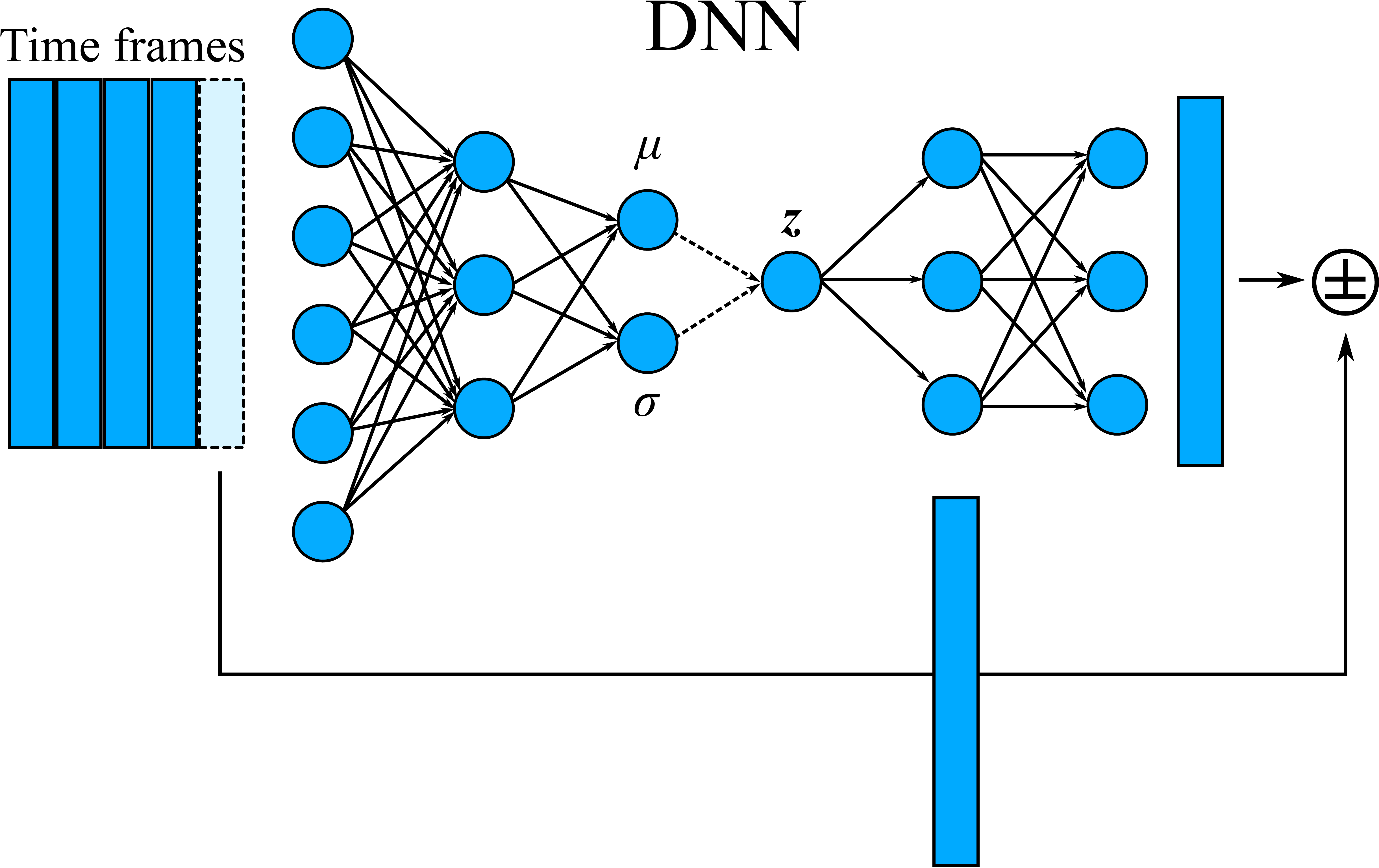}
			\vspace{-1.5em}\subcaption{}\vspace{1em}
			\label{fig:VPDNN}
			\end{minipage}
		\end{tabular}
		\vspace{-2em}
		\caption{Architecture of (a) PDNN and (b) VPDNN}
		\label{fig:PDNN_VPDNN}
	\end{center}
\end{figure}





To solve the issues described above,
our method attempts to only predict the center frame
that is removed from the consecutive frames as the input, which
It can be considered an interpolation of the removed frame.
Thus, we name it ``interpolation DNN (IDNN).''
Figure \ref{fig:IDNN} depicts the proposed architecture of IDNN.
The loss function of IDNN is given as follows:
\begin{align}
  L_{\text{IDNN}}
  = \left\lVert \mathbf{x}_{\frac{n+1}{2}} -
  D( E( \mathbf{x}_{1, \ldots, \frac{n+1}{2}-1, \frac{n+1}{2}+1, \ldots, n} ) )
  \right\rVert_2^2
  ~\text{,}
  \label{eq:loss_func_IDNN}
\end{align}
where $n$ is the sum of the number of the input frames and the output frame.

Given the key assumption that
the detection performance would be improved
by avoiding the difficulty of predicting the edge frames,
an alternative approach named ``prediction DNN (PDNN)'' was also tested
to verify the hypothesis.
Figure \ref{fig:PDNN} shows the architecture of PDNN,
and its loss function is given as follows:
\begin{align}
  L_{\text{PDNN}}
  = \left\lVert \mathbf{x}_{n}-
  D( E( \mathbf{x}_{1,\ldots,n-1} ) )
  \right\rVert_2^2
  ~\text{.}
  \label{eq:loss_func_PDNN}
\end{align}
As illustrated in Figure \ref{fig:PDNN},
consecutive multiple frames are used as an input
and the next frame is predicted as an output.

In addition to the possibility of IDNN described above,
we hypothesize that IDNN has the following merits.
1) It predicts only the center frame
making the number of parameters small,
which enables easier parameter optimization.
2) IDNN can avoid such trivial solutions as an AE
by removing the frame to be predicted from the input
and embedding the spectrotemporal structure of the normal sound.

In both IDNN and PDNN,
the model can be either an AE or a VAE.
Thus, four approaches IDNN with AE/VAE
(named IDNN and VIDNN)
and PDNN with AE/VAE (named PDNN and VPDNN)
were evaluated in this study.
Figure \ref{fig:VIDNN} shows the proposed architecture of VIDNN.
Figure \ref{fig:IDNN} and \ref{fig:VIDNN} show that
IDNN and VIDNN utilize the same input feature vector
and predict the interpolation with each different network.
The concepts of these networks correspond to an AE and a VAE, respectively.
In a similar manner, PDNN and VPDNN predict the next frame
with each different network that corresponds to AE/VAE (see Figure \ref{fig:PDNN_VPDNN}).


\section{Experiment}

\begin{table}[t]
	\begin{center}
	\caption{Summary of dataset}
	\begin{tabular}{@{}l|c@{}}
	Machine types          & Fan, pump, slider, and valve \\
	Data length {[}sec{]}  & 10                           \\
	SNR {[}dB{]}           & -6, 0, 6                     \\
	Sampling rate {[}Hz{]} & 16000                        \\
	\end{tabular}
	\label{tab:MIMII}
	\end{center}
\end{table}

\begin{figure}[t]
	\begin{center}
		\begin{tabular}{cc}
			\begin{minipage}{0.4\hsize}
			\centering
			\includegraphics[width=\columnwidth, clip]{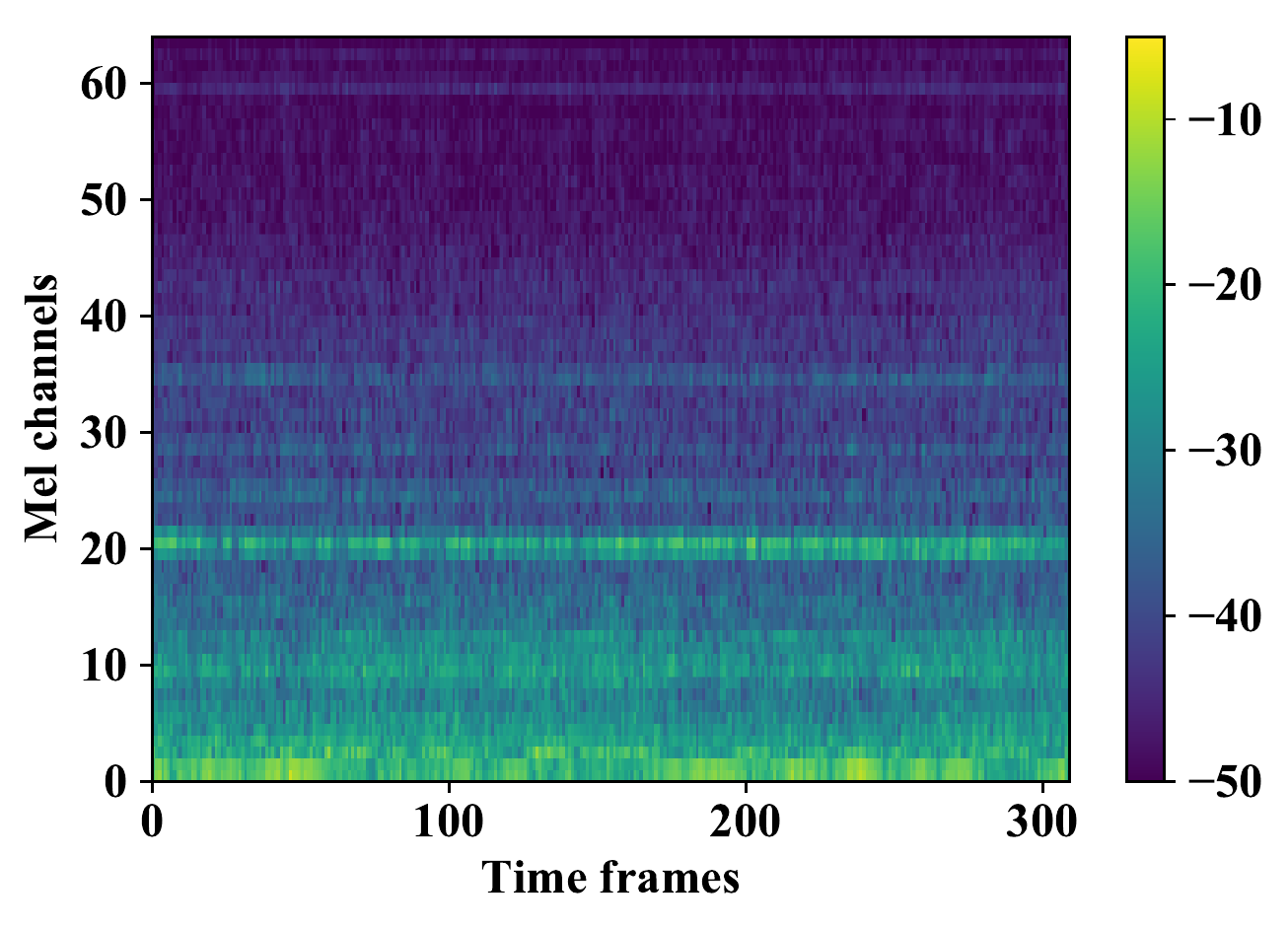}
			\vspace{-1.5em}\subcaption{Fan}\vspace{1em}
			\label{fig:spec_fan}
			\end{minipage}
			&
			\begin{minipage}{0.4\hsize}
			\centering
			\includegraphics[width=\columnwidth, clip]{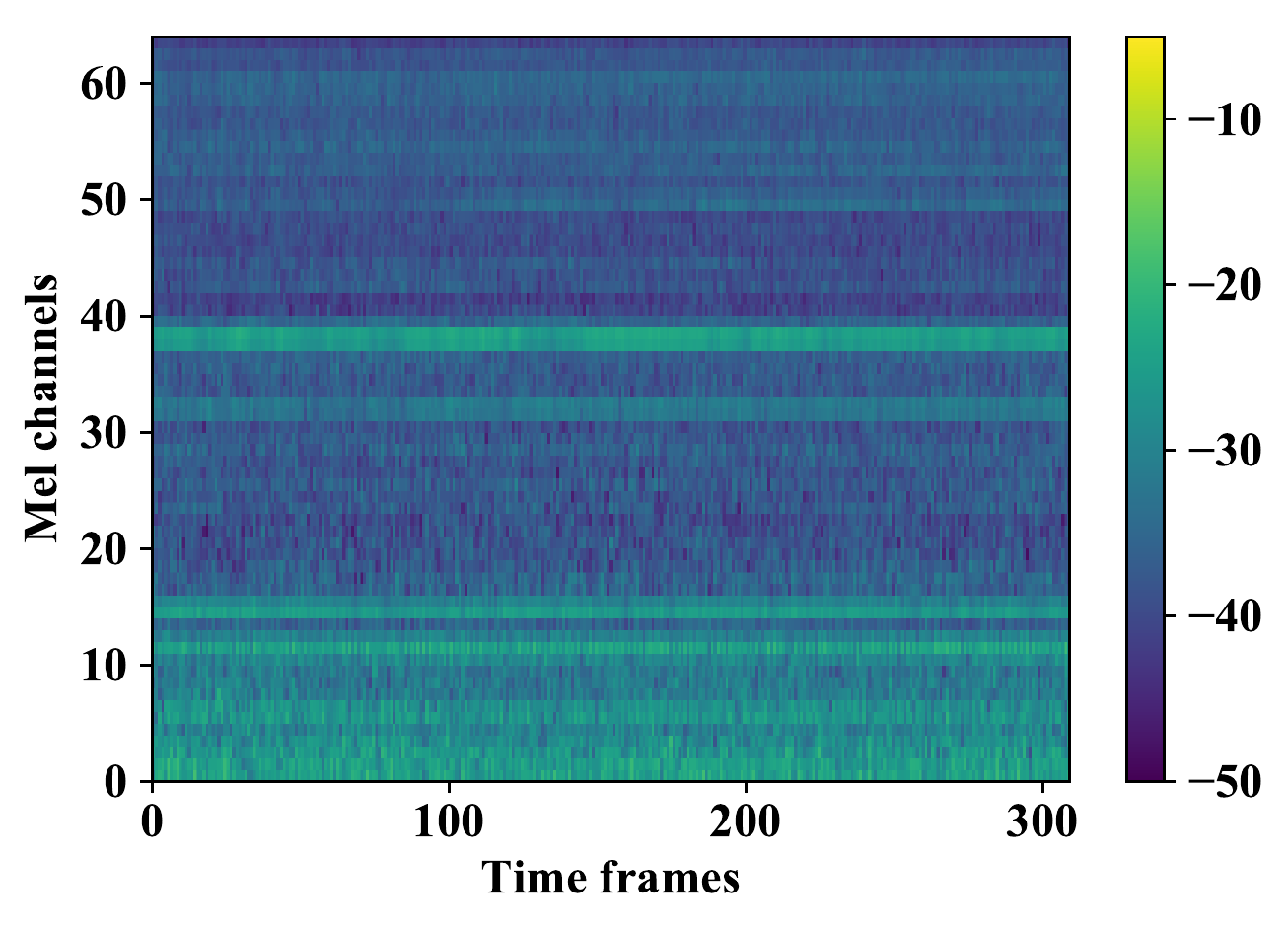}
			\vspace{-1.5em}\subcaption{Pump}\vspace{1em}
			\label{fig:spec_pump}
			\end{minipage}
			\\
			\begin{minipage}{0.4\hsize}
			\centering
			\includegraphics[width=\columnwidth, clip]{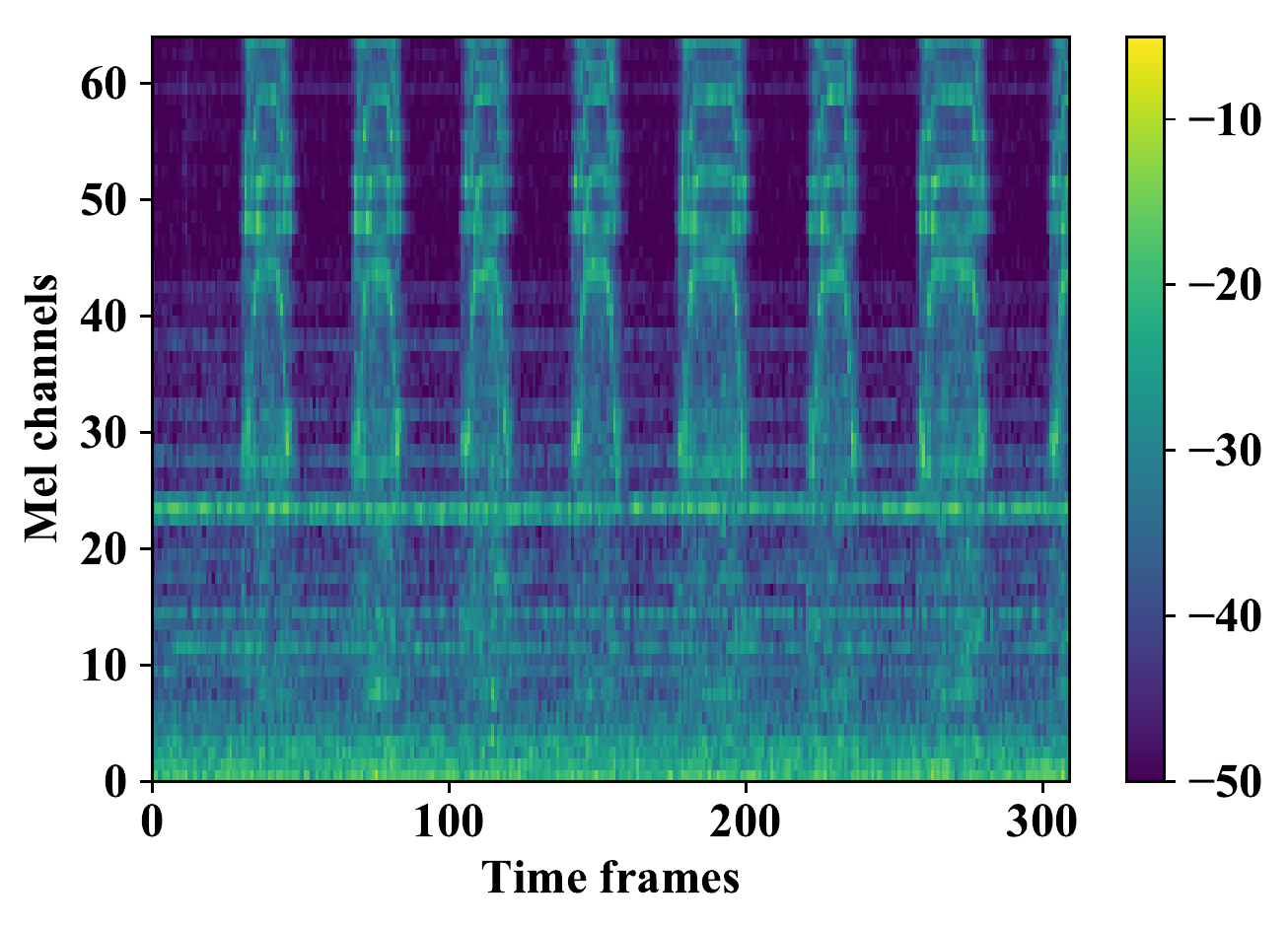}
			\vspace{-1.5em}\subcaption{Slider}\vspace{1em}
			\label{fig:spec_pump}
			\end{minipage}
			&
			\begin{minipage}{0.4\hsize}
			\centering
			\includegraphics[width=\columnwidth, clip]{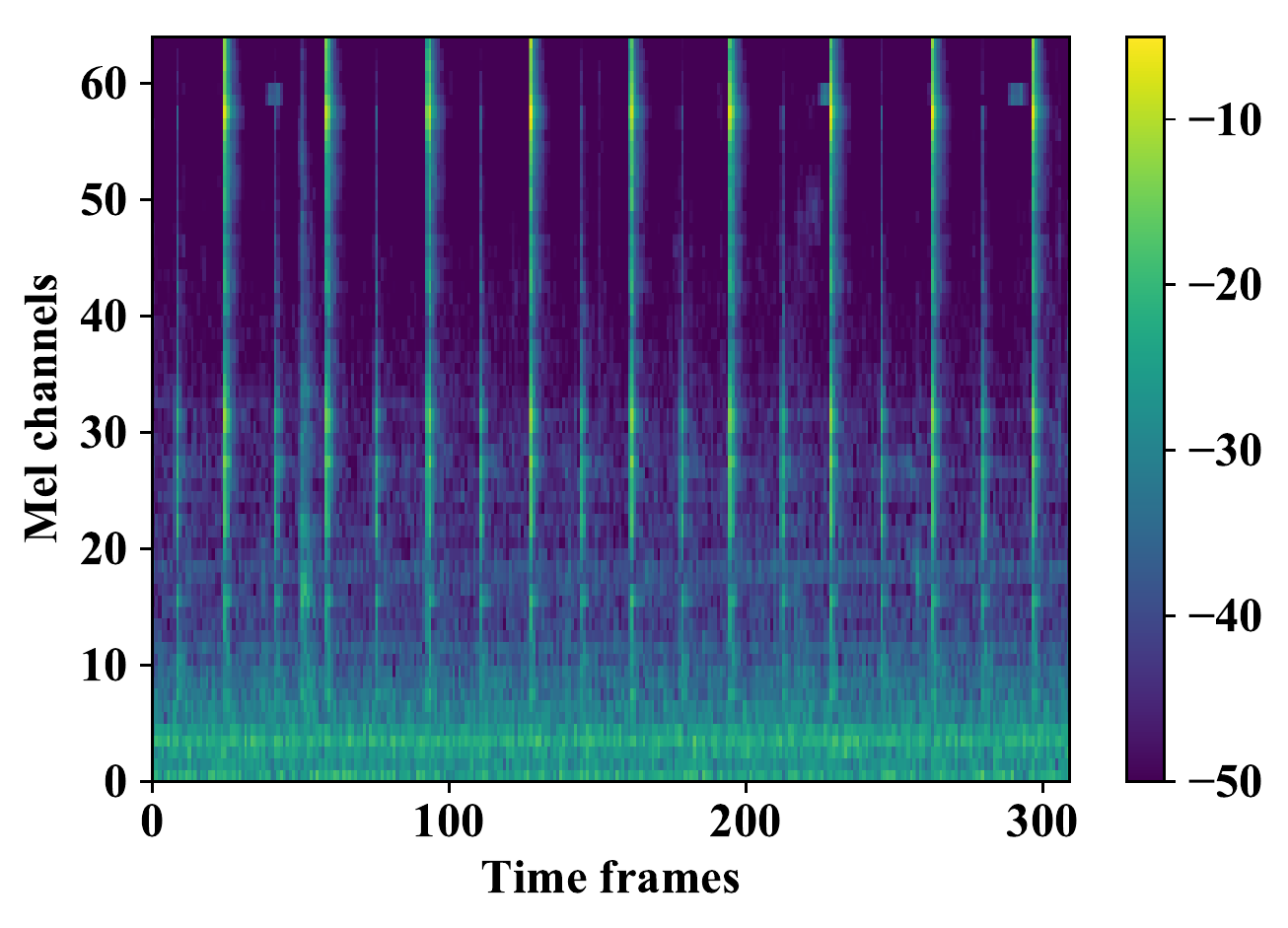}
			\vspace{-1.5em}\subcaption{Valve}\vspace{1em}
			\label{fig:spec_valve}
			\end{minipage}
			\\
		\end{tabular}
		\caption{Examples of log-Mel spectrograms of
		the original sound}
		\label{fig:MIMII_spectrogram}
	\end{center}
\end{figure}
We conducted an experiment using a real-life machinery sound dataset \cite{MIMII}
to evaluate
the performance of our approach.
Table \ref{tab:MIMII} summarizes the dataset.
There were a total of 24,490 normal sound segments
and 5,620 anomalous sound segments.
Each machine type consists of seven individual machines.

For our IDNN, PDNN, and the conventional approach,
an AE and a VAE were trained for each machine type.
A log-Mel spectrogram was used as an input feature.
To calculate the Mel spectrogram,
the frame size was set to 1024, the hop size was set to 512, and the number of Mel filter banks was set to 64.
For the conventional approaches,
five frames were concatenated and used as an input feature vector,
and the same number of frames were reconstructed as an output.
For our approaches,
four frames were used as an input,
and one frame was interpolated/predicted as an output.

The autoencoder network structure for the
experiment is summarized as follows:
The encoder network E($\cdot$)
comprises FC(Input, 64, ReLU), FC(64, 32, ReLU),
and FC(32, 16, ReLU);
the decoder network D($\cdot$) incorporates FC(16, 32, ReLU)
FC(32, 64, ReLU), and
FC(64, Output, none),
where FC($a, b, f$) represents a fully-connected layer
with input neurons $a$, an output layer $b$,
and activation function $f$, respectively \cite{Jarrett2009}.
The network was trained with an Adam optimization technique \cite{Kingma2015Adam}.
The weight coefficient $w$ in Eq.\ \ref{eq:loss_func_VAE}
was empirically optimized to 0.1, 0.01, and 0.01
for the VAE, VIDNN, and VPDNN, respectively.
The performance was evaluated based on
the area under the curve (AUC) of the receiver operating characteristic,
and the calculation was iterated
three times for each individual machine.

\begin{figure}[t]
	\begin{center}
		\includegraphics[width=.75\columnwidth, clip]{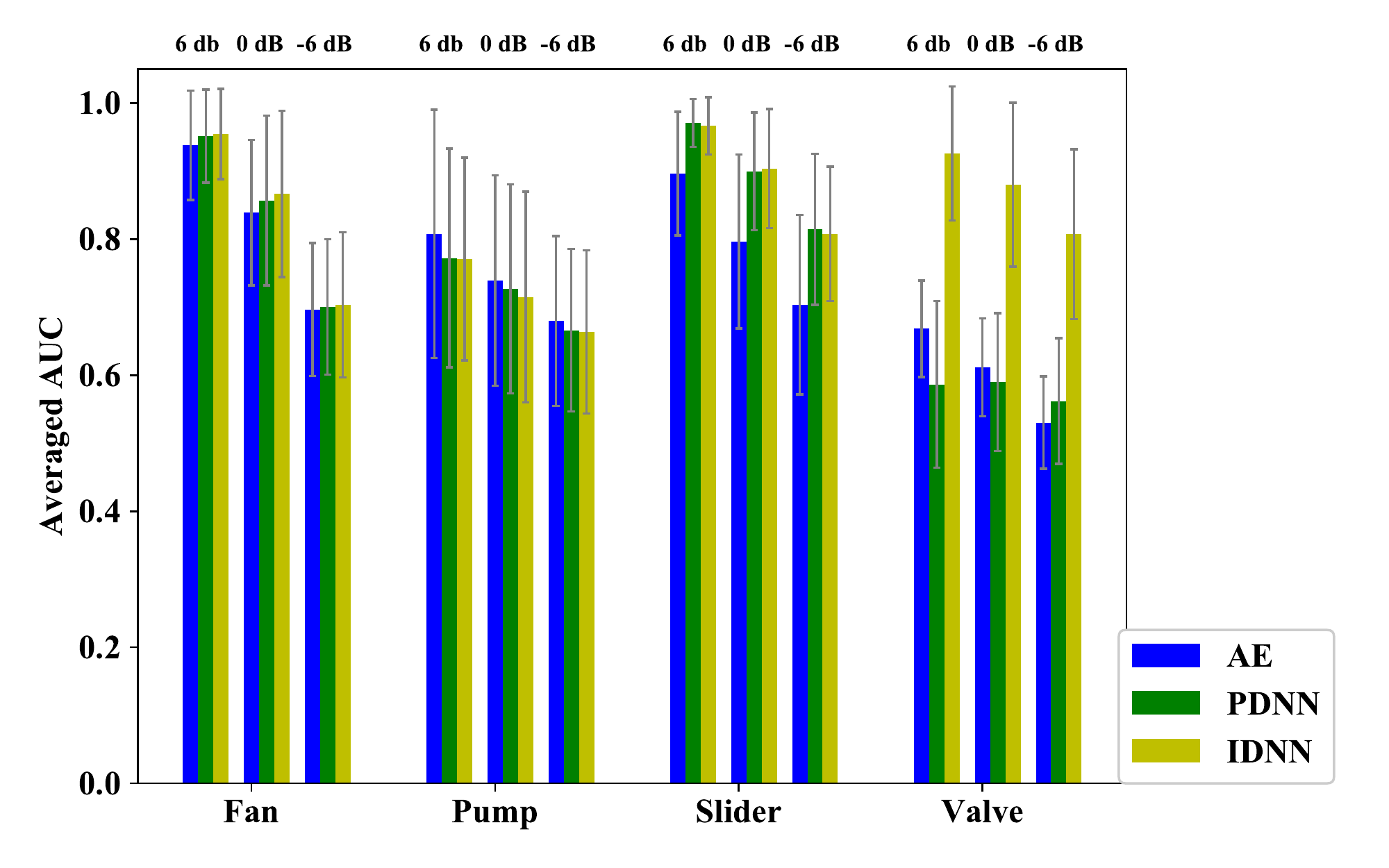}
		\caption{Averaged AUC of the AE, IDNN, and PDNN}
		\label{fig:AUC_AE}
	\end{center}
\end{figure}

\begin{figure}[t]
	\begin{center}
		\includegraphics[width=.75\columnwidth, clip]{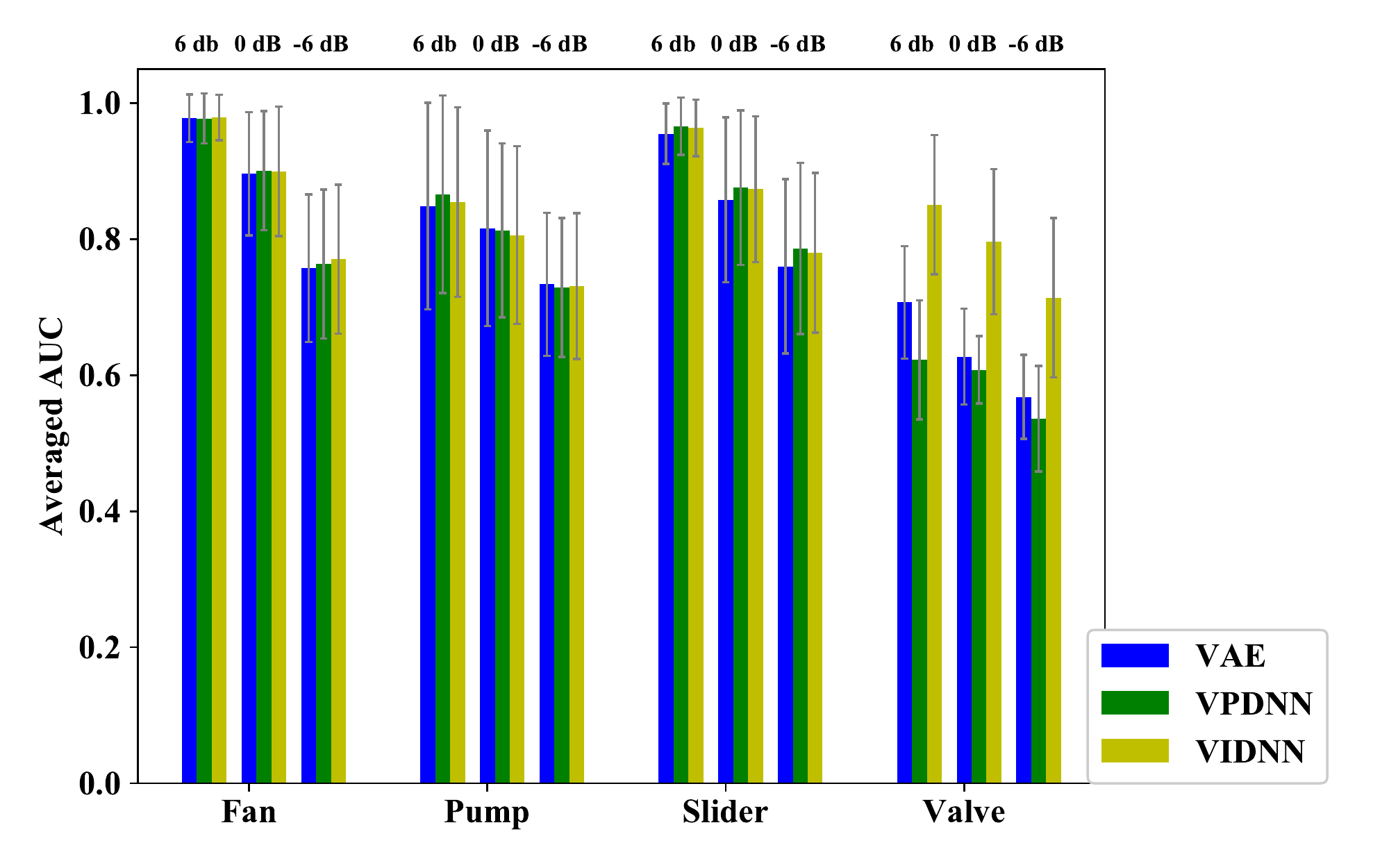}
		\caption{Averaged AUC of the VAE, VIDNN, and VPDNN}
		\label{fig:AUC_VAE}
	\end{center}
\end{figure}

Figure \ref{fig:AUC_AE} 
shows the results of averaged AUC with the AE, IDNN, and PDNN.
Figure \ref{fig:AUC_VAE} 
shows the results of averaged AUC with the VAE, VIDNN, and VPDNN.
As depicted in Figure \ref{fig:AUC_AE},
the proposed IDNN showed significantly higher AUC
compared to the AE and PDNN with the valve sound.
With the slider sound,
IDNN and PDNN both showed higher AUC than the AE.
On the other hand, 
IDNN and PDNN and the conventional approach
performed similarly with the fan and the pump sound.
Meanwhile, as depicted in Figure \ref{fig:AUC_VAE},
our VIDNN and VPDNN performed similarly
to the conventional VAE
except for the valve sound
where VIDNN outperformed the VAE and VPDNN.
A similar trend can be seen regardless of SNR
in Figs.\ \ref{fig:AUC_AE} and \ref{fig:AUC_VAE}

\begin{figure}[t]
	\begin{center}
		\begin{tabular}{cc}
			\begin{minipage}{0.45\hsize}
			\centering
			\includegraphics[width=\columnwidth, clip]{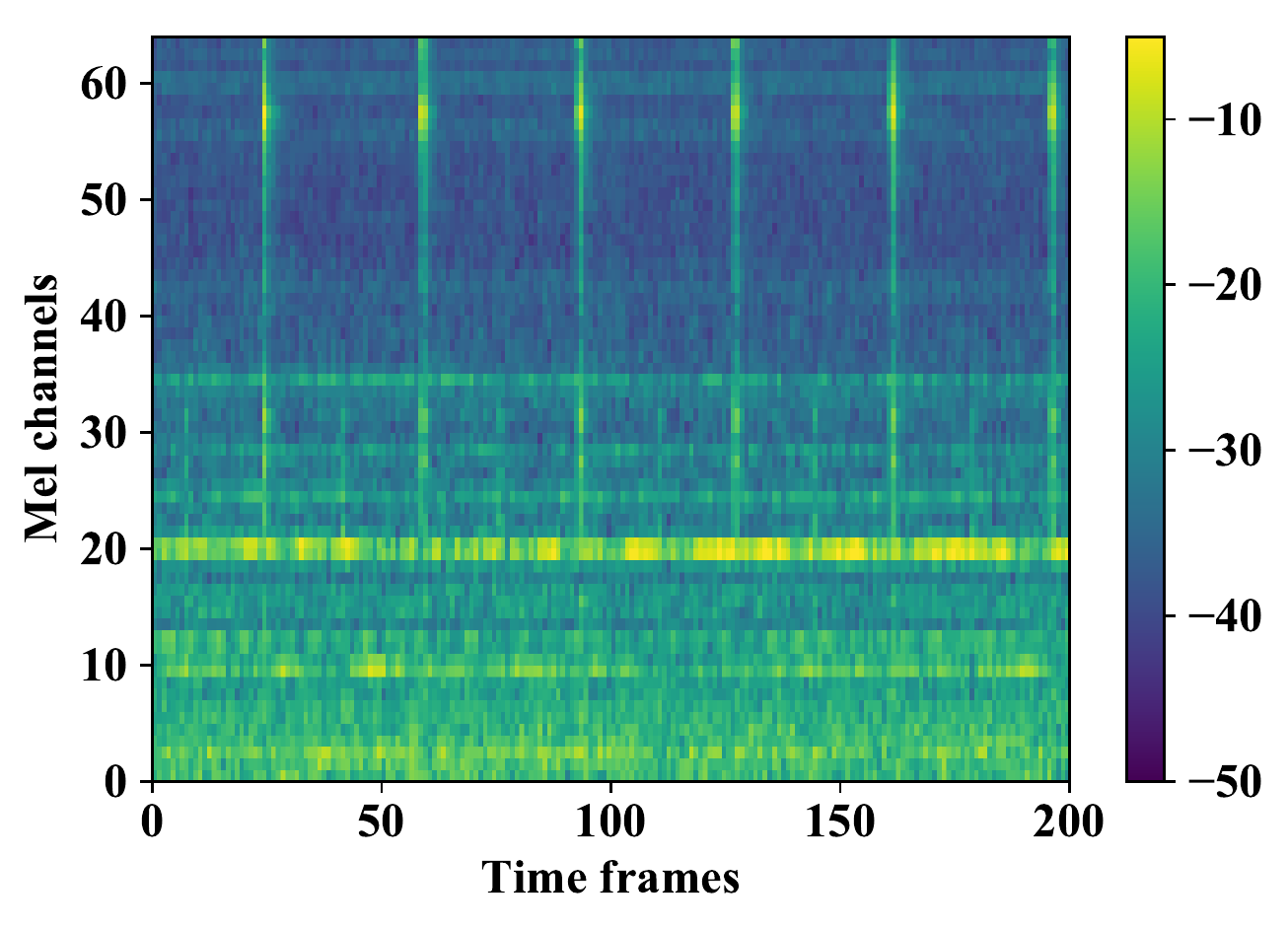}
			\vspace{-1.5em}\subcaption{Input}\vspace{1em}
			\label{fig:input}
			\end{minipage}
			&
			\\
			\begin{minipage}{0.45\hsize}
			\centering
			\includegraphics[width=\columnwidth, clip]{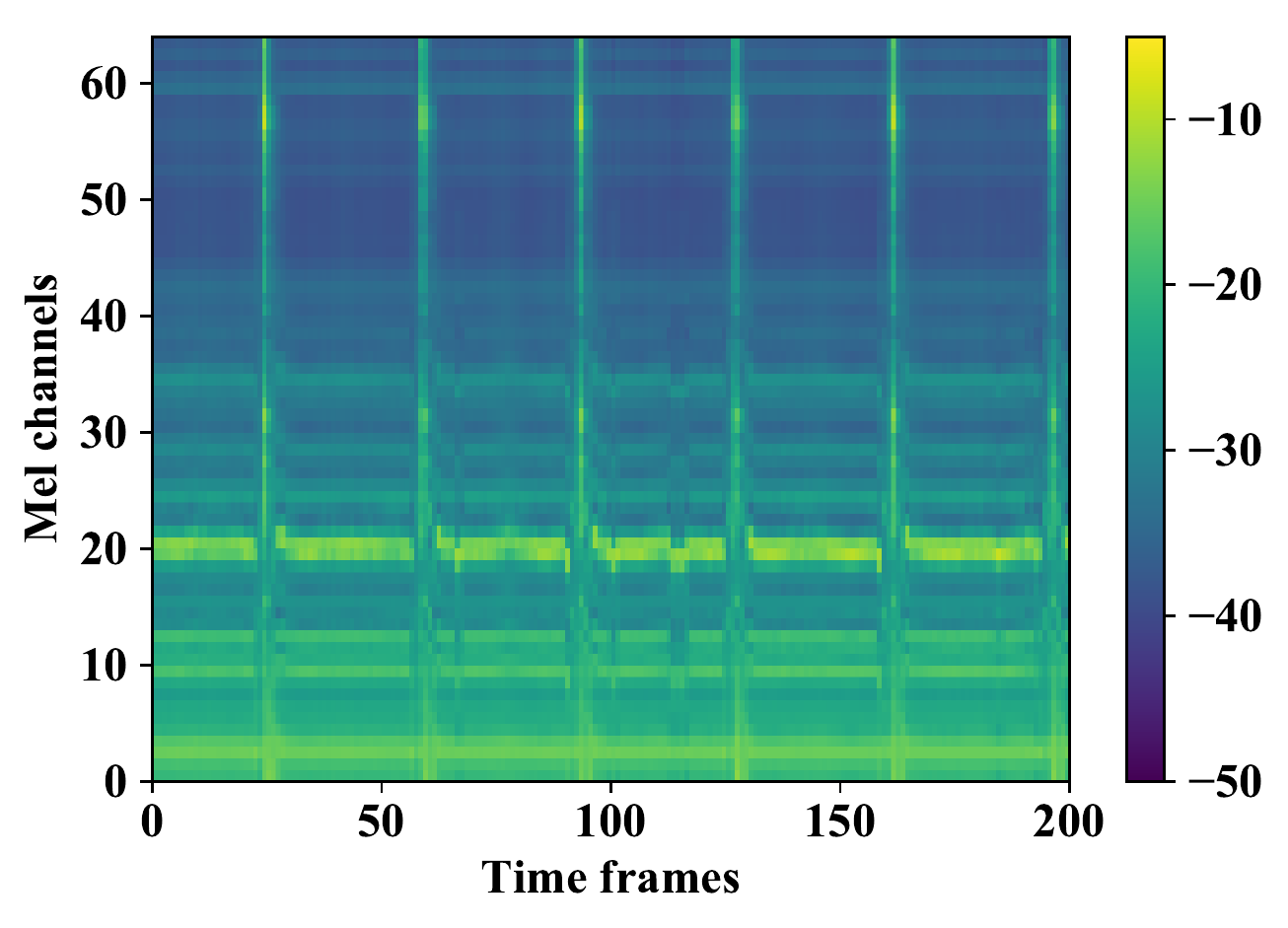}
			\vspace{-1.5em}\subcaption{Output of AE}\vspace{1em}
			\label{fig:out_AE}
			\end{minipage}
			&
			\begin{minipage}{0.45\hsize}
			\centering
			\includegraphics[width=\columnwidth, clip]{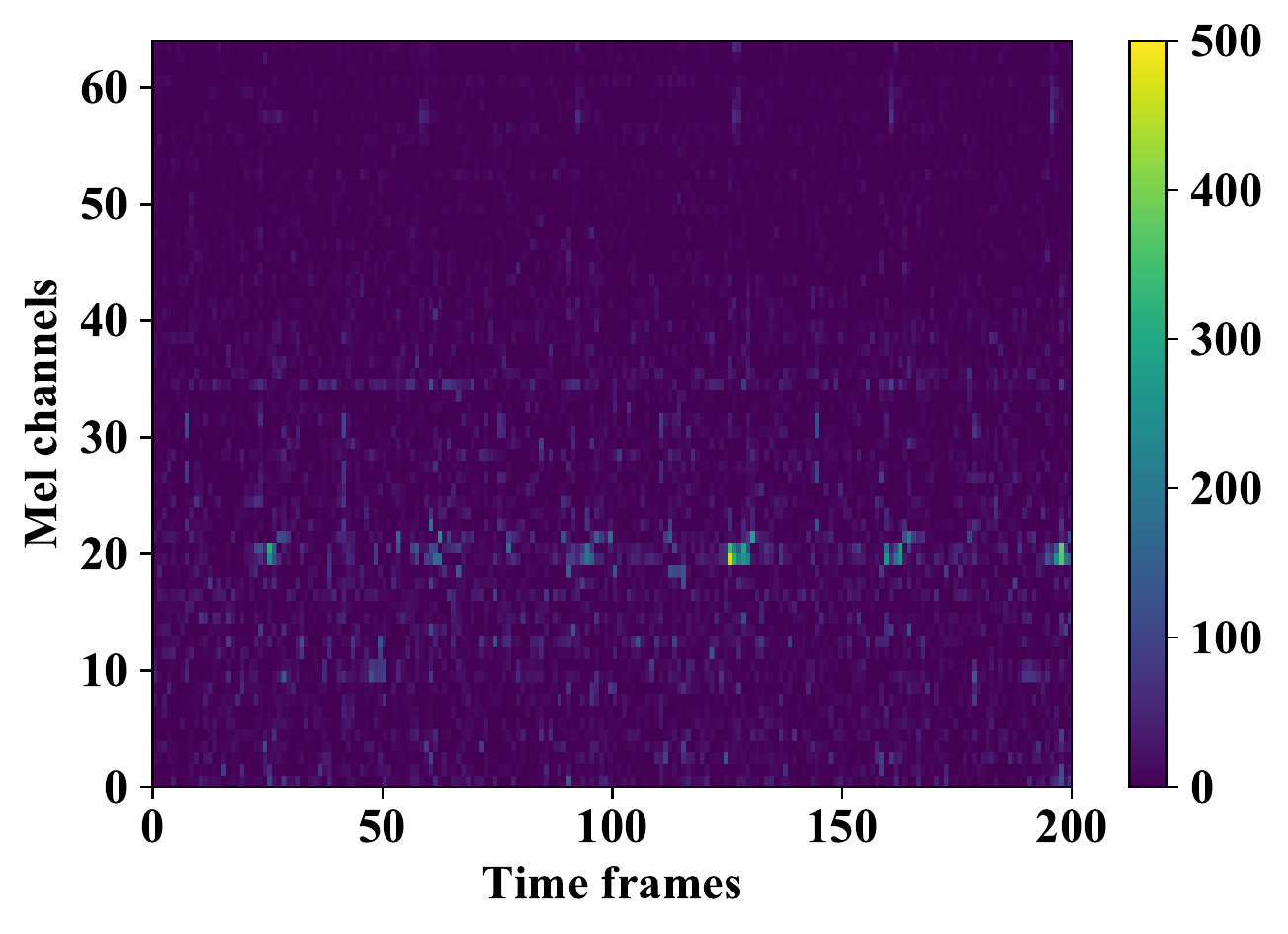}
			\vspace{-1.5em}\subcaption{Error of AE}\vspace{1em}
			\label{fig:err_AE}
			\end{minipage}
			\\
			\begin{minipage}{0.45\hsize}
			\centering
			\includegraphics[width=\columnwidth, clip]{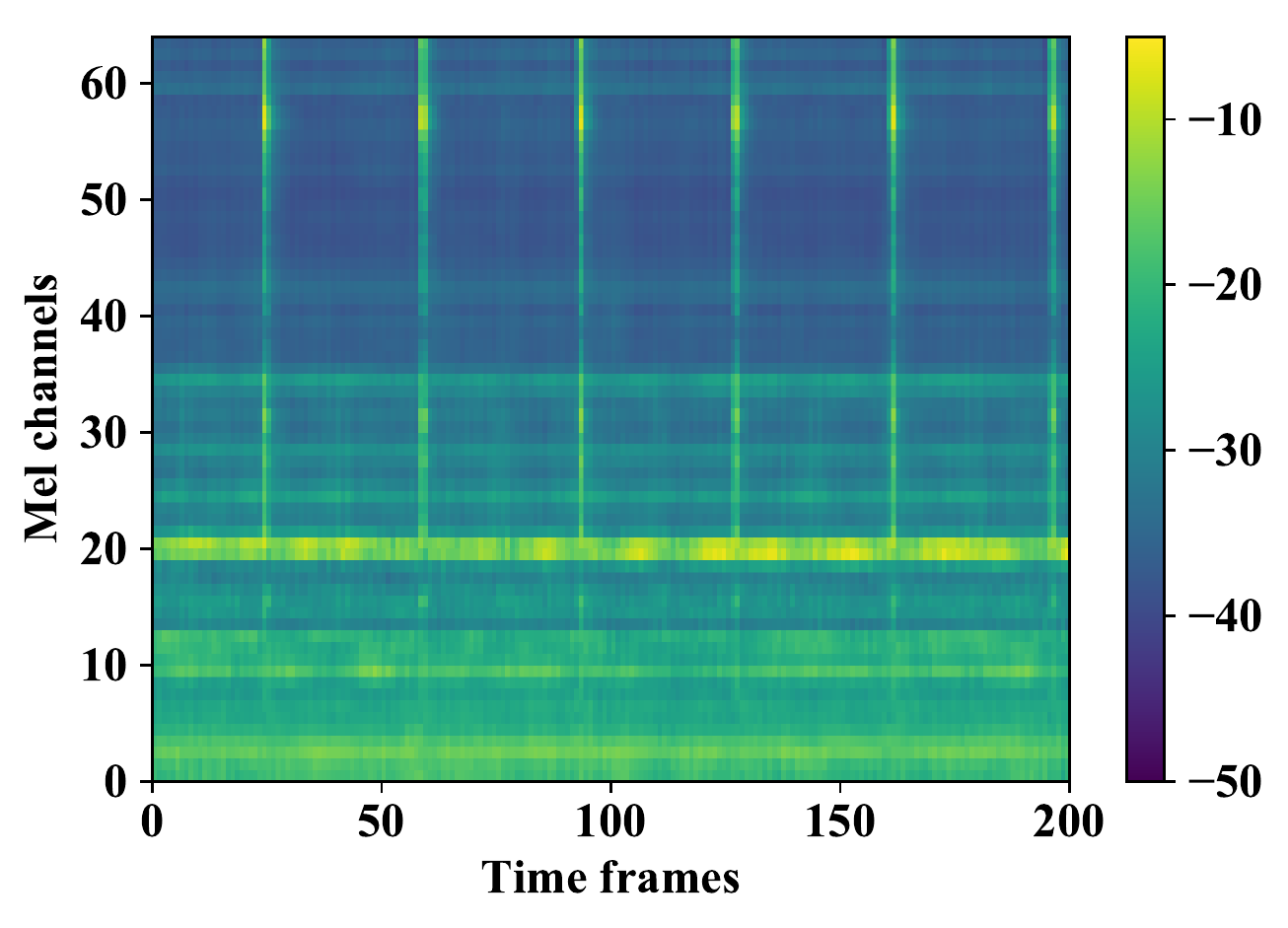}
			\vspace{-1.5em}\subcaption{Output of IDNN}\vspace{1em}
			\label{fig:out_IDNN}
			\end{minipage}
			&
			\begin{minipage}{0.45\hsize}
			\centering
			\includegraphics[width=\columnwidth, clip]{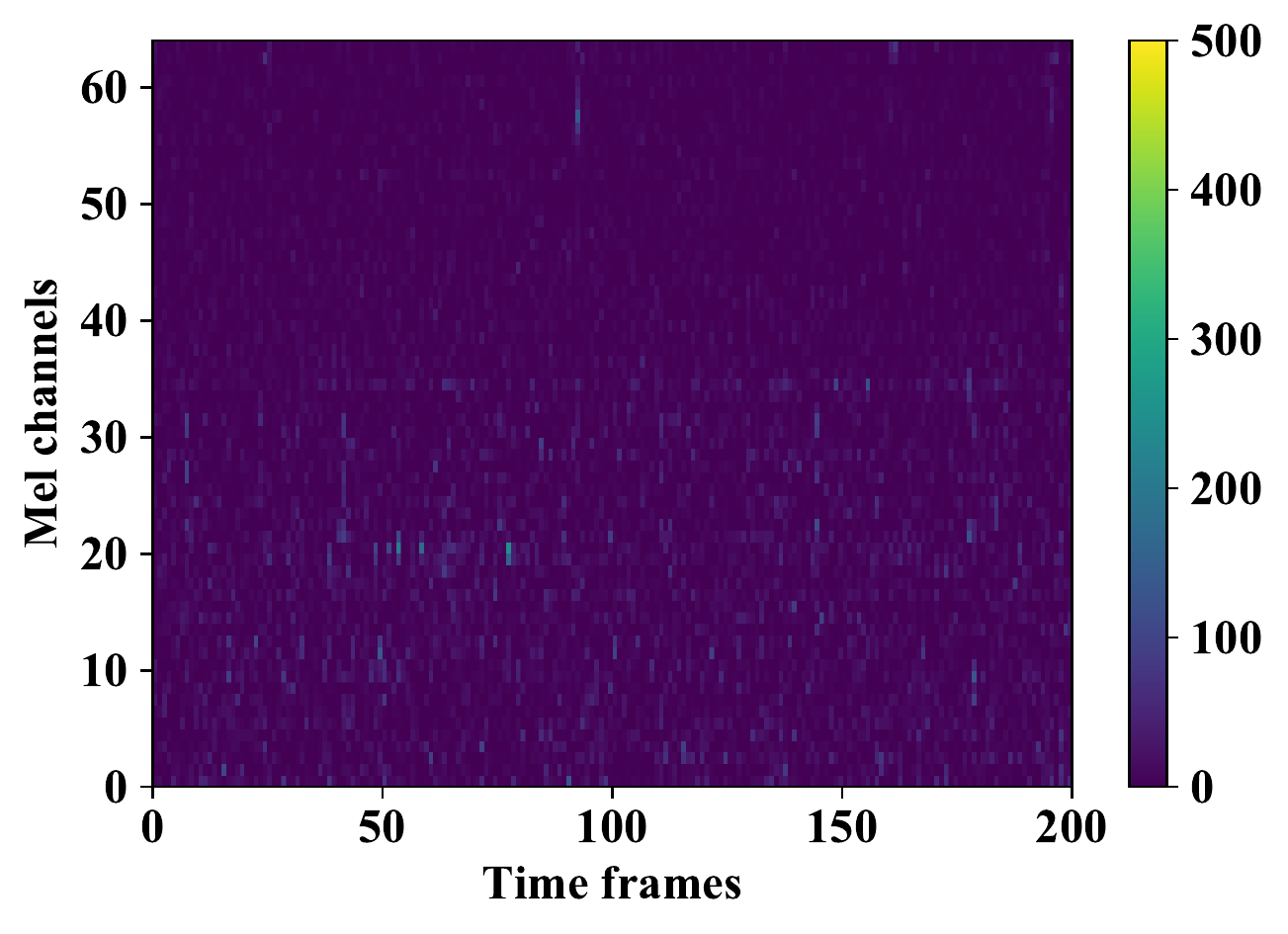}
			\vspace{-1.5em}\subcaption{Error of IDNN}\vspace{1em}
			\label{fig:err_IDNN}
			\end{minipage}
			\\
			\begin{minipage}{0.45\hsize}
			\centering
			\includegraphics[width=\columnwidth, clip]{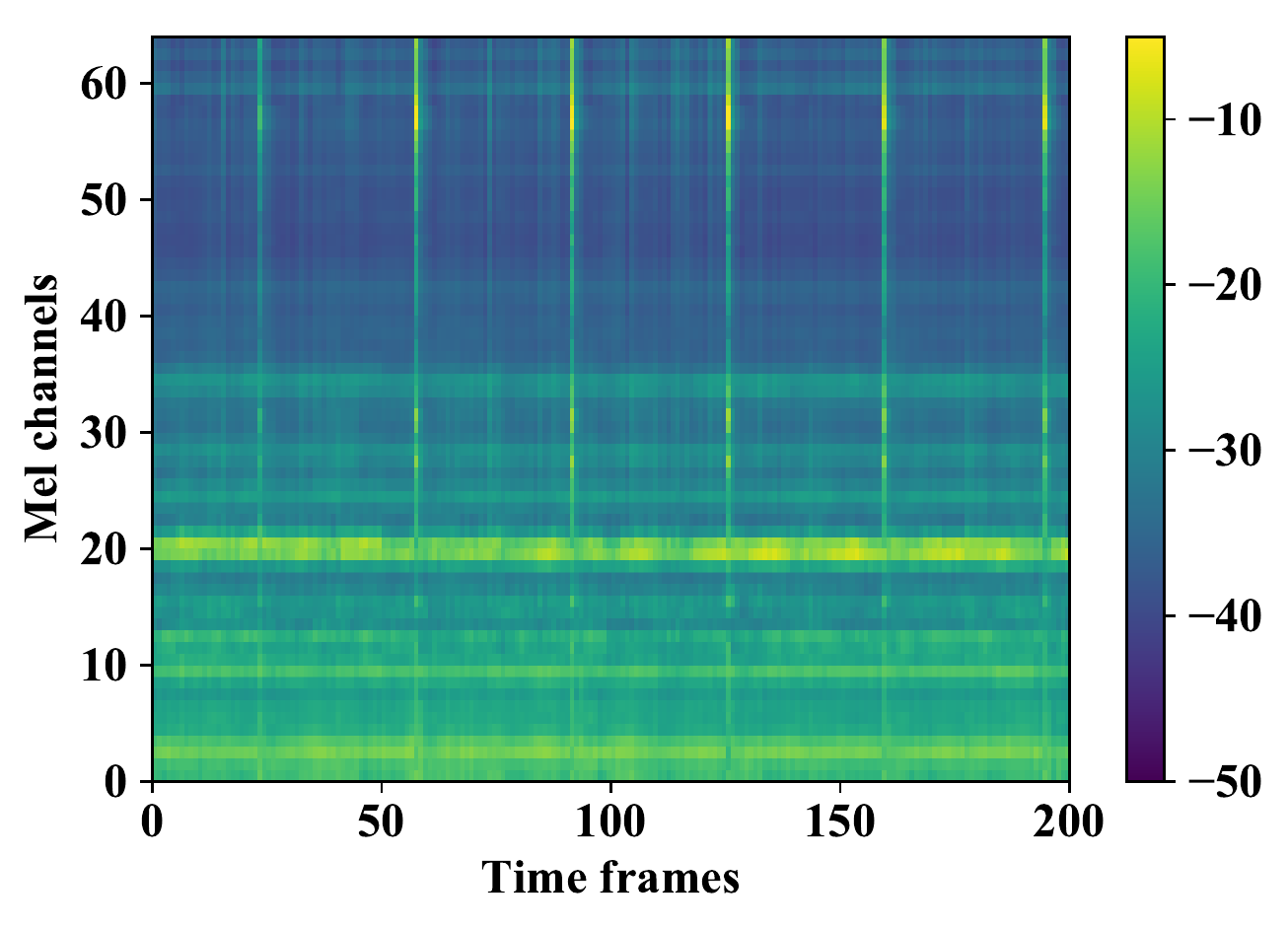}
			\vspace{-1.5em}\subcaption{Output of PDNN}\vspace{1em}
			\label{fig:out_PDNN}
			\end{minipage}
			&
			\begin{minipage}{0.45\hsize}
			\centering
			\includegraphics[width=\columnwidth, clip]{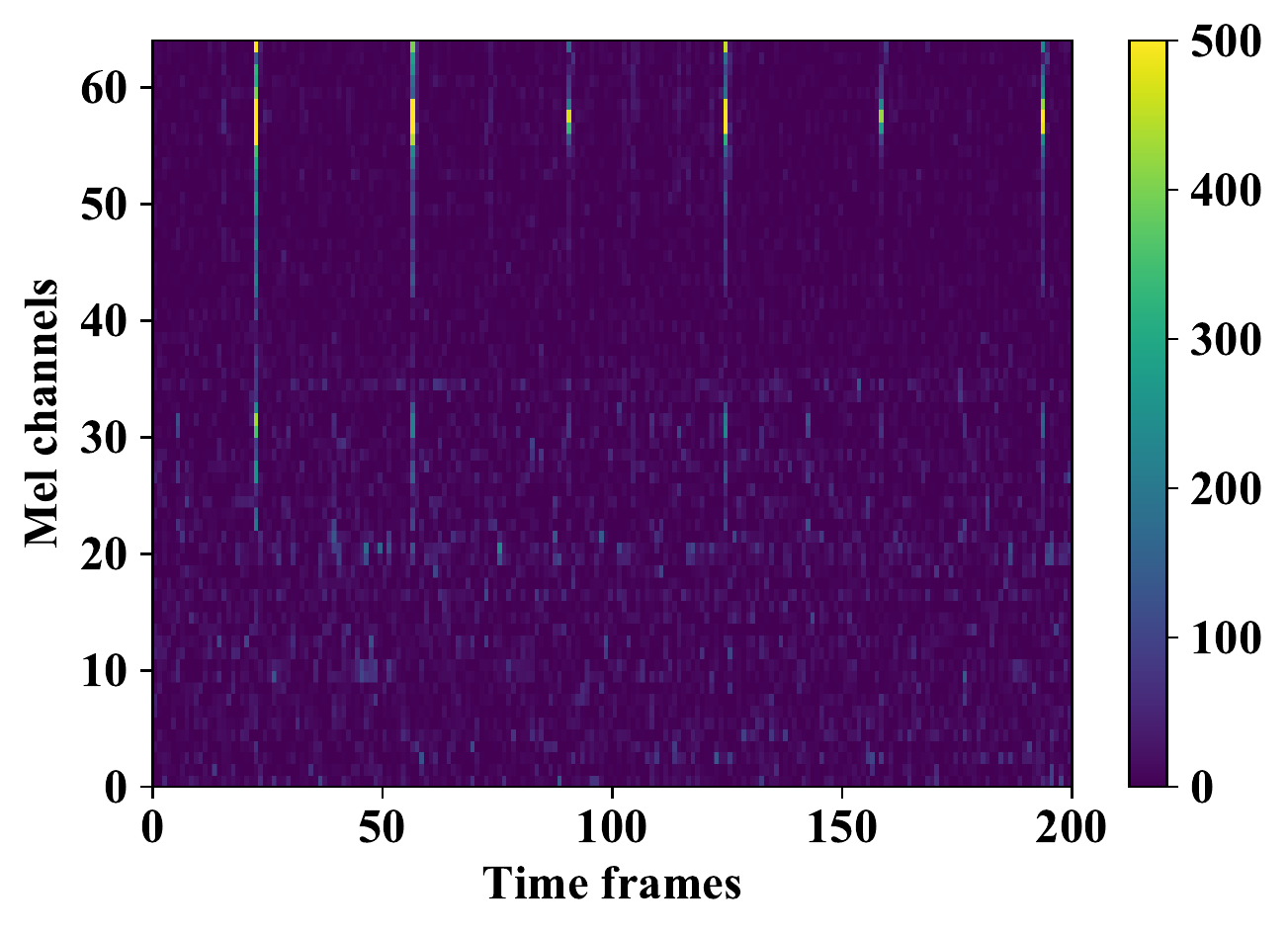}
			\vspace{-1.5em}\subcaption{Error of PDNN}\vspace{1em}
			\label{fig:err_PDNN}
			\end{minipage}
			\\
		\end{tabular}
		\caption{Examples of restoration of the normal valve sound}
		\label{fig:rest_normal}
	\end{center}
\end{figure}

\begin{figure}[t]
	\begin{center}
		\begin{tabular}{cc}
			\begin{minipage}{0.45\hsize}
			\centering
			\includegraphics[width=\columnwidth, clip]{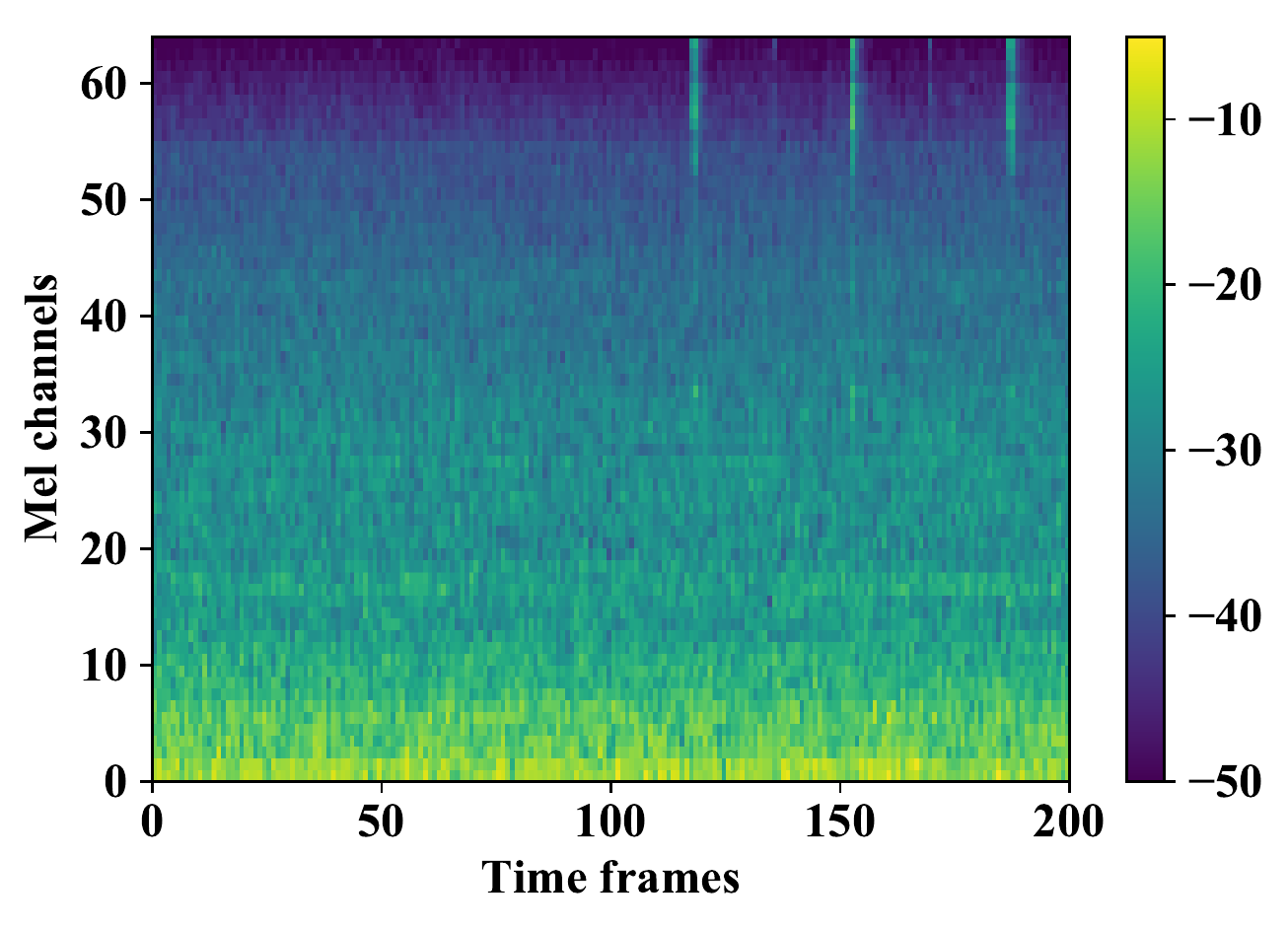}
			\vspace{-1.5em}\subcaption{Input}\vspace{1em}
			\label{fig:input_abnoraml}
			\end{minipage}
			&
			\\
			\begin{minipage}{0.45\hsize}
			\centering
			\includegraphics[width=\columnwidth, clip]{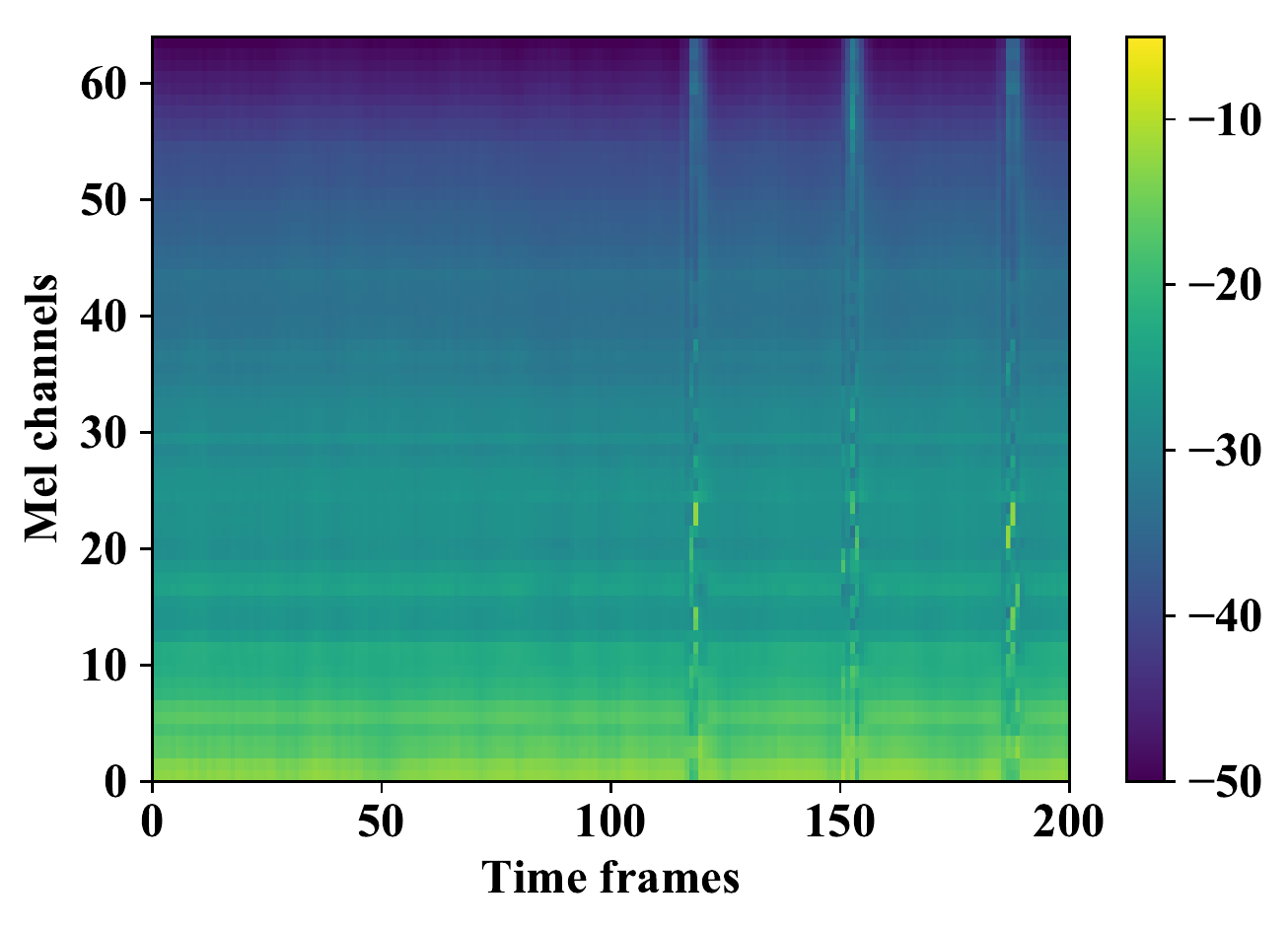}
			\vspace{-1.5em}\subcaption{Output of AE}\vspace{1em}
			\label{fig:out_abnoraml_AE}
			\end{minipage}
			&
			\begin{minipage}{0.45\hsize}
			\centering
			\includegraphics[width=\columnwidth, clip]{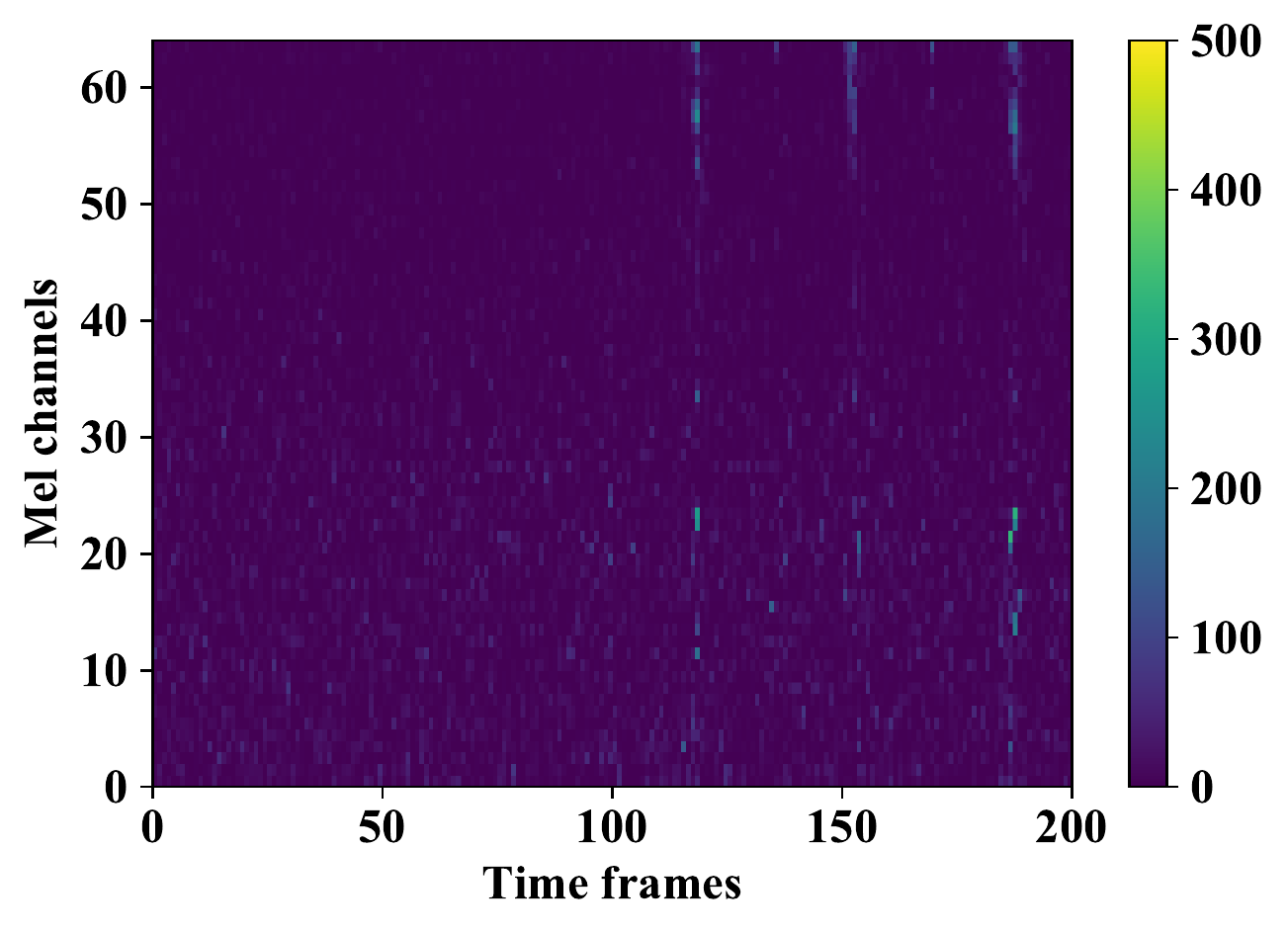}
			\vspace{-1.5em}\subcaption{Error of AE}\vspace{1em}
			\label{fig:err_abnoraml_AE}
			\end{minipage}
			\\
			\begin{minipage}{0.45\hsize}
			\centering
			\includegraphics[width=\columnwidth, clip]{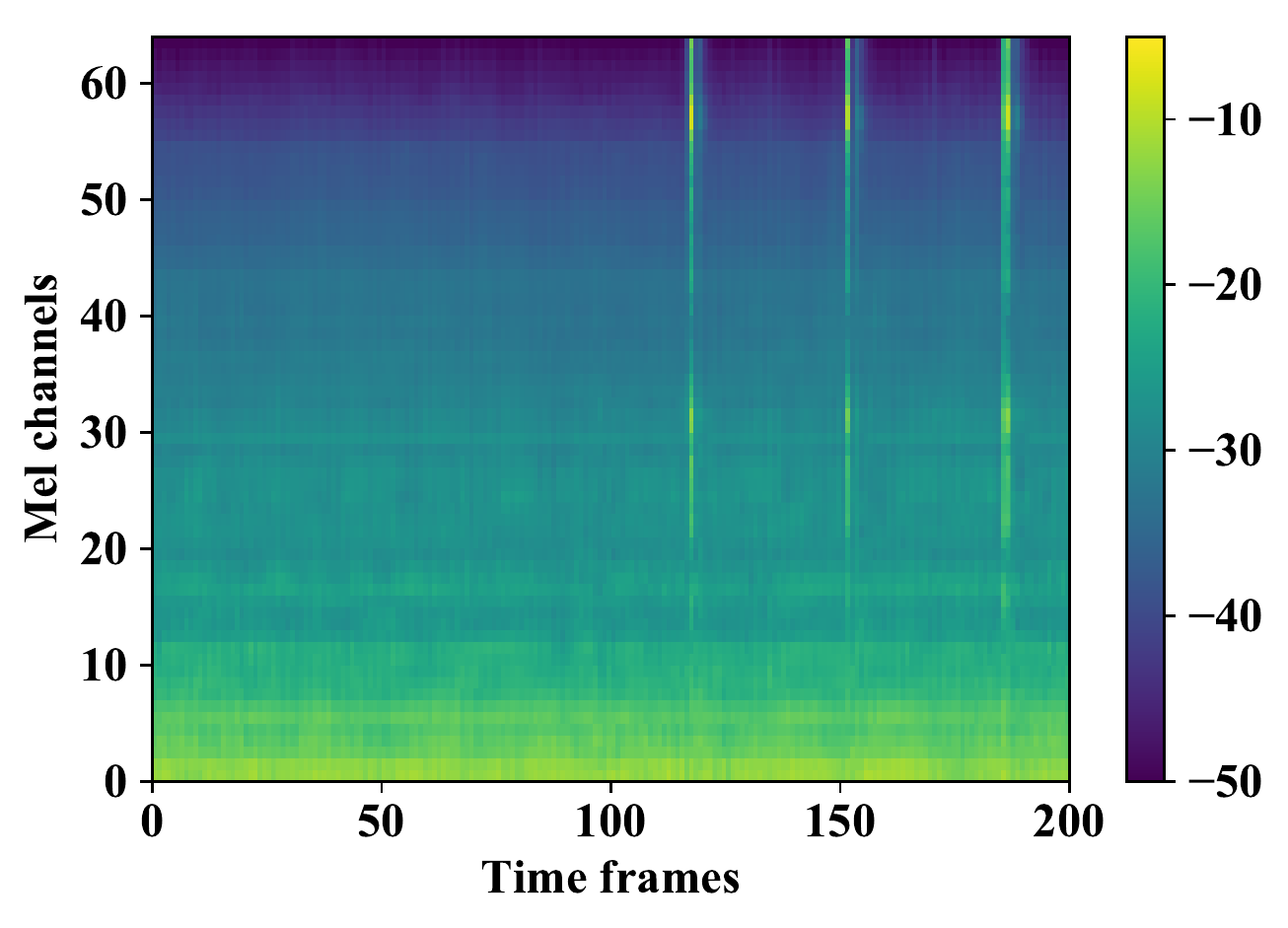}
			\vspace{-1.5em}\subcaption{Output of IDNN}\vspace{1em}
			\label{fig:out_abnoraml_IDNN}
			\end{minipage}
			&
			\begin{minipage}{0.45\hsize}
			\centering
			\includegraphics[width=\columnwidth, clip]{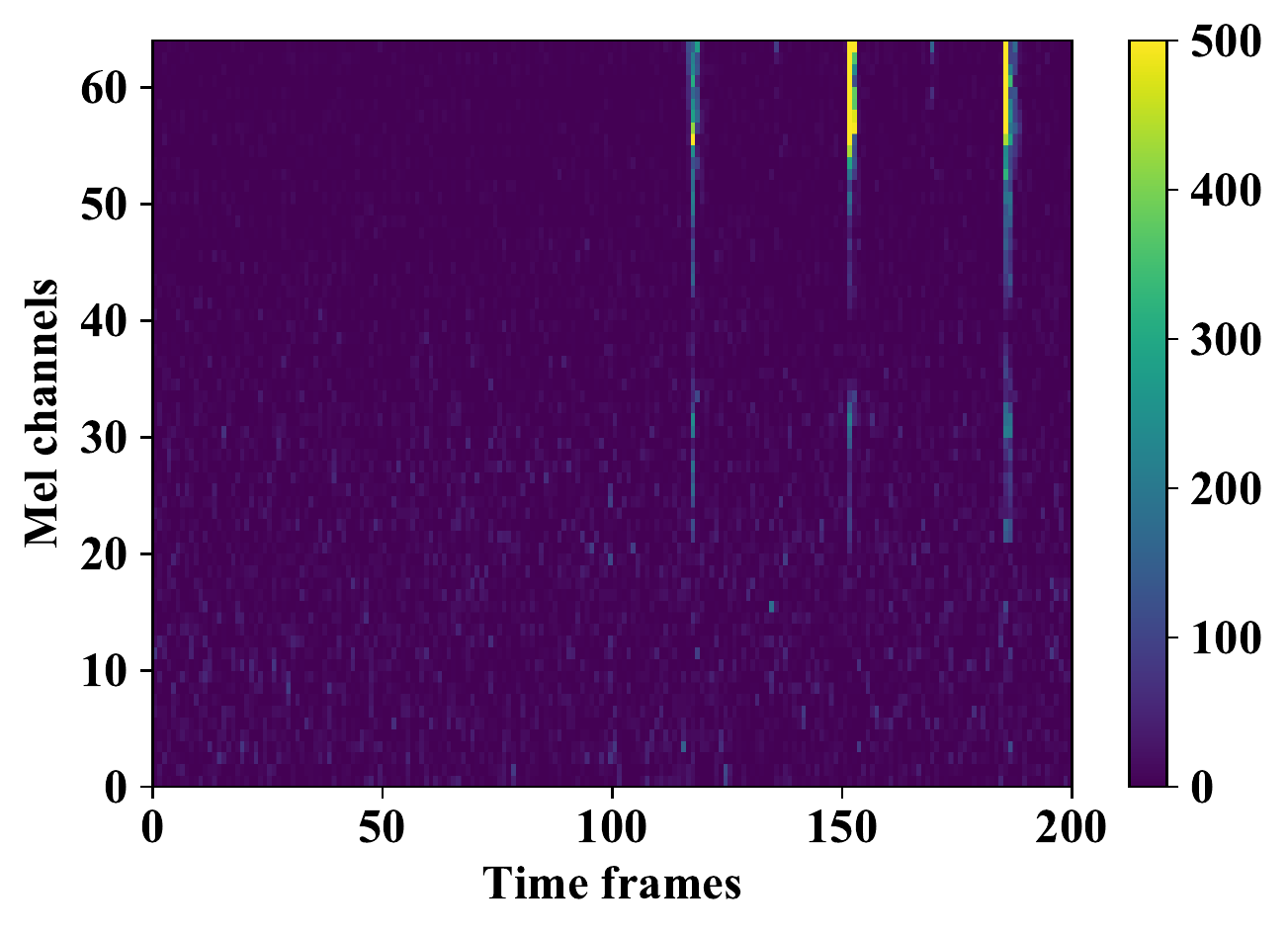}
			\vspace{-1.5em}\subcaption{Error of IDNN}\vspace{1em}
			\label{fig:err_abnoraml_IDNN}
			\end{minipage}
			\\
			\begin{minipage}{0.45\hsize}
			\centering
			\includegraphics[width=\columnwidth, clip]{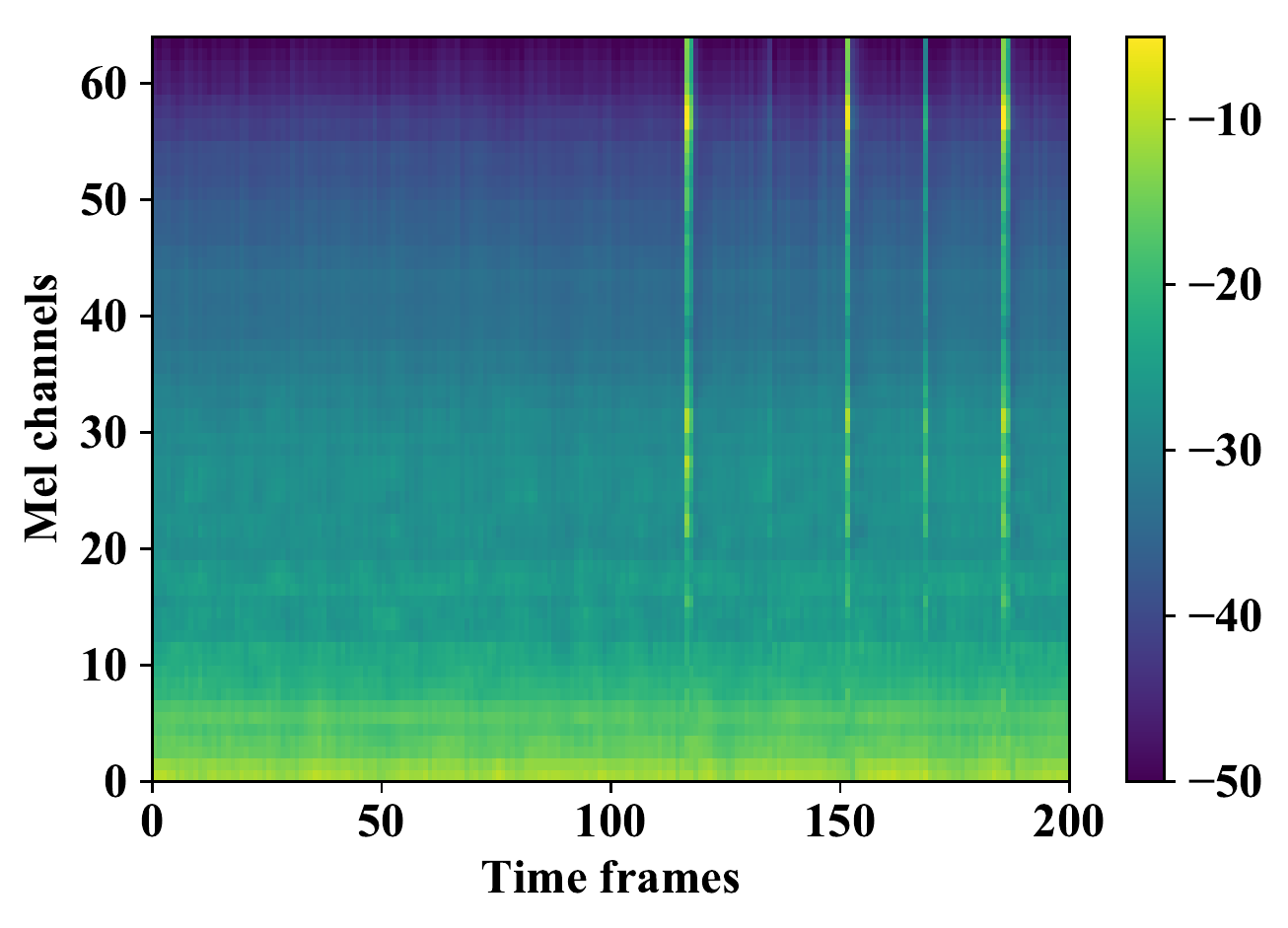}
			\vspace{-1.5em}\subcaption{Output of PDNN}\vspace{1em}
			\label{fig:out_abnormal_PDNN}
			\end{minipage}
			&
			\begin{minipage}{0.45\hsize}
			\centering
			\includegraphics[width=\columnwidth, clip]{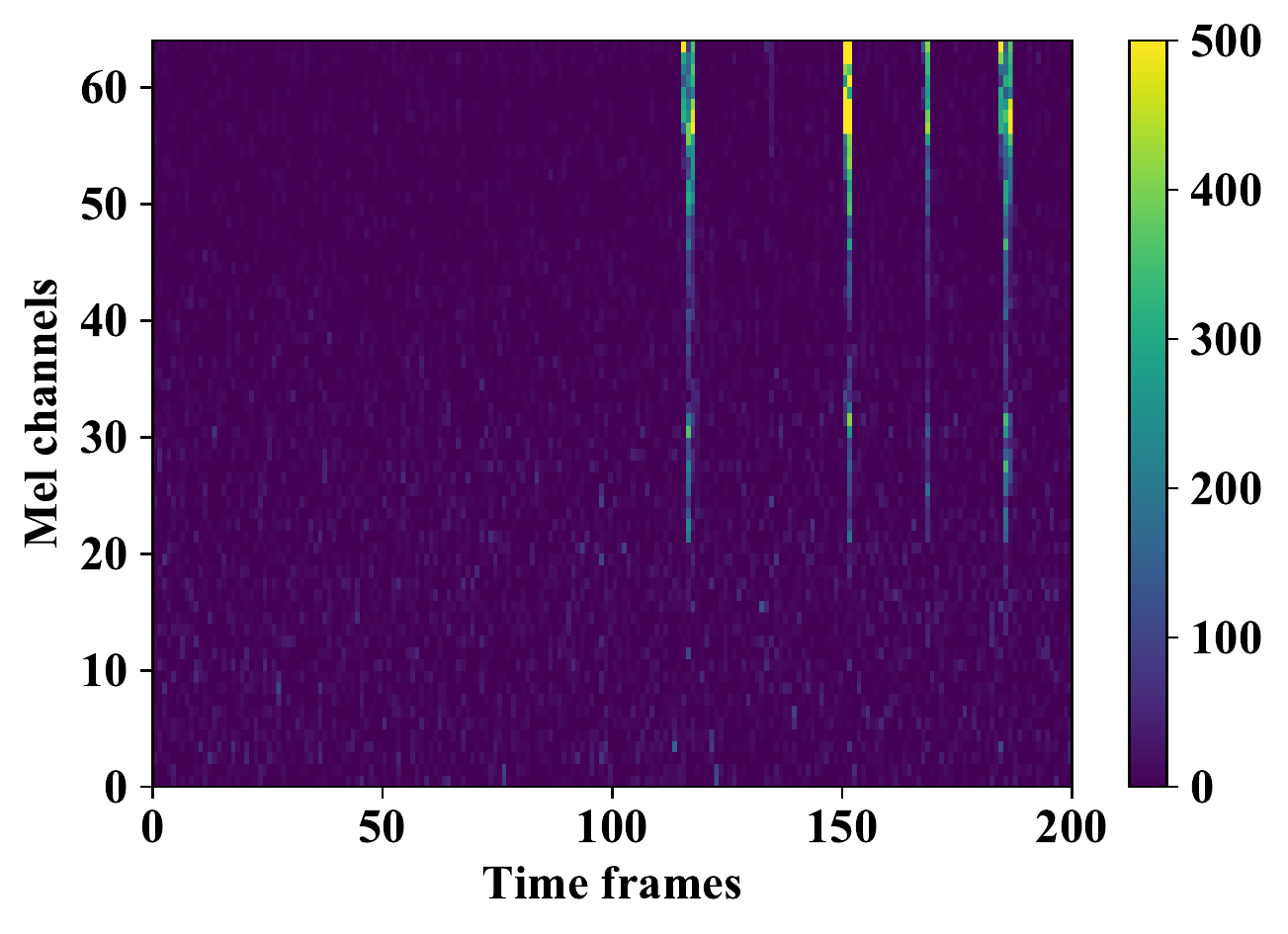}
			\vspace{-1.5em}\subcaption{Error of PDNN}\vspace{1em}
			\label{fig:err_abnormal_PDNN}
			\end{minipage}
			\\
		\end{tabular}
		\caption{Examples of restoration of the anomalous valve sound}
		\label{fig:rest_abnormal}
	\end{center}
\end{figure}

As Figure \ref{fig:MIMII_spectrogram} shows,
non-stationarity can be seen in the valve and the slider sound,
where the proposed IDNN outperformed the conventional approach.
For the following discussions,
the performances were compared based on an example of the valve sound.

Figures \ref{fig:rest_normal} and \ref{fig:rest_abnormal} show
the restored output
for the normal and abnormal sound of the valve, respectively.
As shown in Figure \ref{fig:rest_normal},
both the AE and IDNN removed noises well
and showed small errors with the normal sound.
Meanwhile, in the case of
the anomalous sound (see Figure \ref{fig:rest_abnormal}),
the error (i.e., anomaly score) of IDNN was properly large
while the AE showed a smaller error than that of IDNN.
As Figure \ref{fig:rest_abnormal} shows,
the spectrogram reconstructed by IDNN was similar to
that of the normal valve,
indicating that IDNN was accurately trained for non-stationarity.
In contrast, the error of PDNN was large without regarding normality,
indicating that predicting the edge frame is more difficult than
interpolating the center frame.
Additionally, as shown in Figure \ref{fig:AUC_AE},
IDNN and PDNN showed similar performance with the slider sound,
while IDNN performed much better than PDNN with the valve sound.
In terms of the characteristics of the sound,
the sound changes of the valve were shorter
than the slider (see Figure \ref{fig:MIMII_spectrogram}), indicating that
the sound change in a shorter duration made
predicting the edge frame more difficult,
and IDNN can be more robust in such a situation.


\section{Conclusion}
\label{sec:conclusion}
We proposed an approach to anomalous sound detection
that employs an interpolation error of AE/VAE as an anomaly score
that avoids the difficulty of predicting the edge frame.
Experimental results showed that our approach
outperformed conventional approaches
for the non-stationary sound in particular.
In the study,
the number of input frames and the output were set to four and one, respectively.
Further studies are needed in order to assess how those parameters can affect the detection rate.



\bibliographystyle{IEEEbib}
\bibliography{refs}

\end{document}